\documentclass[12 pt,fleqn]{article}
\usepackage{amsmath,a4,latexsym,epsfig}
\newcommand\mathbox[2]{\makebox[#1][l]{$\displaystyle #2$}}
\renewcommand{\L}{{\cal L}}
\newcommand{\lambdabar}{{\overline\lambda}}
\newcommand{\sigmabar}{{\overline\sigma}}
\newcommand{\psibar}{{\overline\psi}}
\newcommand{\epsilonbar}{{\overline\epsilon}}
\newcommand{\thetabar}{{\overline\theta}}
\newcommand{\chibar}{{\overline\chi}}
\newcommand{\Psibar}{{\overline\Psi}}
\newcommand{\alphadot}{{\dot\alpha}}
\newcommand{\betadot}{{\dot\beta}}
\newcommand{\gammadot}{{\dot\gamma}}
\newcommand{\deltadot}{{\dot\delta}}
\newcommand{\intx}{\int d^4x}
\def\dg#1{\frac{\delta\Gamma}{\delta#1}}
\def\dF#1{\frac{\delta{\cal F}}{\delta#1}}
\def\dfunc#1#2{\frac{\delta^{#1}}{\delta#2}}
\def\dpartial#1#2{\frac{\partial^{#1}}{\partial#2}}
\def\pslash#1{{\setbox0=\hbox{$#1$}
  \rlap{\ifdim\wd0>.7em\kern.22\wd0\else\kern.1\wd0\fi /}#1}}
\def\psl{\pslash p}

\def\qsl{\pslash q}
\def\Dsl{\pslash D}

\begin{document}



\begin{titlepage}

\begin{flushright}
KA--TP--8--1999\\
BN--TH--99--11\\
{\tt hep-ph/9907393}\\
\end{flushright}

\begin{center}
{\Large\bf{
        Renormalization and Symmetry Conditions\\[1ex]
        in Supersymmetric QED}}
\\
\vspace{8ex}
{\large W. Hollik$^a$,
        E. Kraus$^b$,
        D. St{\"o}ckinger$^a$
{\renewcommand{\thefootnote}{\fnsymbol{footnote}}
\footnote{E-mail addresses:\\
                hollik@particle.physik.uni-karlsruhe.de,\\
                kraus@theo11.physik.uni-bonn.de,\\
                ds@particle.physik.uni-karlsruhe.de.}} 
\\
\vspace{2ex}
{\small\em $^a$ Institut f{\"u}r Theoretische Physik, 
              Universit{\"a}t Karlsruhe,\\
              D--76128 Karlsruhe, Germany\\
}
\vspace{.5ex}
{\small\em               $^b$ Physikalisches Institut,
              Universit{\"a}t Bonn,\\
              Nu{\ss}allee 12, D--53115 Bonn, Germany\\}}
\vspace{2ex}
\end{center}

{\small
 {\bf Abstract}
 \newline\newline
 For supersymmetric gauge theories a consistent
regularization scheme that preserves supersymmetry and gauge
invariance is not known.
In this article we tackle this problem for
supersymmetric QED within the framework of algebraic renormalization.
For practical calculations, a non-invariant regularization scheme 
may be used together with counterterms from all
power-counting renormalizable interactions.
From the Slavnov--Taylor identity, expressing gauge
invariance, supersymmetry and translational invariance,
simple symmetry conditions are derived
that are important in a twofold respect:
they establish exact relations between 
physical quantities  
that are valid to all orders, and 
they provide a powerful tool for the practical
determination of the counterterms. 
We perform concrete one-loop calculations
in dimensional regularization, where supersymmetry is
spoiled  at the regularized level, and show how the counterterms
necessary to restore supersymmetry can be read off easily. 
In addition, a specific example is given how 
the supersymmetry transformations in one-loop order 
are modified by non-local terms.

}

\end{titlepage}


\newpage
\section{Introduction}

In phenomenological studies of the electroweak standard model (SM)
and its extensions it is crucial to take into account radiative
corrections. Comparing theoretical predictions with experimental
precision data provides tests and comparisons of the models at
the level of their quantum structure. In particular, as far as
collider energies are too low to produce Higgs or e.g.~supersymmetric 
particles, this is the only way to obtain information about such
heavy sectors.

The calculation of these radiative corrections involves a technical
problem. The loop integrals are in general divergent and need 
regularization. But this procedure can break essential symmetries of
the underlying theory, such as gauge invariance or supersymmetry. The two
most important regularization schemes for the SM and its
supersymmetric extensions are dimensional regularization (DReg) \cite{DReg,BM77}
and dimensional reduction (DRed) \cite{Siegel79}, the difference being
that in the latter case only the momenta are treated $D$-dimensional
whereas the vector fields and $\gamma^\mu$ matrices are not.

As already noted by the inventor, DRed is inconsistent
\cite{Siegel80}: it is possible to derive the equation
$0=D(D-1)(D-2)(D-3)(D-4)$ in contradiction to regularization at
$D\ne4$. On the other hand, DReg breaks supersymmetry 
whereas DRed was designed to preserve
supersymmetry\cite{Siegel79,CJN80}. There are many modifications of
both schemes; for example, in \cite{ACV81} a version of DRed was
suggested which is mathematically consistent but not
supersymmetric. In fact, no consistent regularization scheme is known
that simultaneously preserves supersymmetry and gauge
invariance for supersymmetric gauge theories. A similar problem arises
in chiral gauge theories like the standard model. 

For practical calculations an invariant scheme is desirable. So in
most phenomenological applications requiring supersymmetric
calculations schemes such as DRed are used
together with arguments that the inconsistencies do not show up in
the actual cases \cite{JaJo97}. But these arguments have a
restricted range of validity, and it is not yet clear if and how they
may be applied to calculations beyond one loop in the SM
and its supersymmetric extensions \cite{Weig99}. 

In this article we pursue the opposite way: Instead of searching for an
invariant regularization we advocate the use of arbitrary
regularization schemes and define the finite (renormalized) Green
functions by the basic symmetries, as it is proposed by the abstract
approach of algebraic renormalization. (For an introduction to
algebraic renormalization see ref.~\cite{PiSo}.) 

From an abstract point of view, the question of the existence of a
symmetry-preserving scheme is irrelevant. The theory is defined by
symmetry requirements that should be satisfied after renormalization. 
There are two equivalent ways to achieve that. The first way is to
use an invariant scheme keeping the symmetries manifest. In this case,
only those counterterms are necessary for renormalization that themselves
preserve the symmetries. These are usually just the ones obtained by
multiplicative renormalization of the parameters and fields in the
Lagrangian of the theory. The second way is to use a non-invariant
scheme and to compensate the corresponding symmetry breaking by
appropriate non-invariant counterterms. Although less obvious, this
possibility was noted in many milestones of renormalization theory,
e.g.~in \cite{Bogoliubov,tHooftRen,BRS}. 
Generally, by using a non-invariant scheme a precise definition of the
symmetries one requires from the renormalized theory is mandatory. In
order to establish these symmetries one has to allow for all possible
counterterms, restricted only by hermiticity, Lorentz invariance and
power counting renormalizability, but not by further symmetries.  

Of course, if there exists no scheme that keeps the symmetries
manifest there could be anomalies making it impossible to restore the
symmetries by adjusting the counterterms. But the absence of
anomalies, too, may be proven without any recurrence to a particular
regularization, only using algebraic properties of the symmetry
requirements \cite{BRS}.

The first algebraic analysis of renormalizability of  supersymmetric gauge
theories was performed in the superspace formalism \cite{PiSi84}.
For phenomenological applications it is preferable to use the component 
formulation of supersymmetric gauge theories in the Wess--Zumino
gauge, where the unphysical fields are eliminated by the
supersymmetric gauge transformation. Finding a well defined identity
expressing the symmetry content of supersymmetric gauge theories in the
Wess--Zumino gauge is not easy since there the supersymmetry algebra
does not close but also includes gauge transformations  (see
eq.~(\ref{SusyAlgebra})). In particular, a separate treatment of gauge
invariance and supersymmetry seems impossible --- one would need
infinitely many sources and renormalizability could not be proven 
\cite{BrMa85}. The solution of this problem was found in \cite{HLW90,White92a}
combining ideas of Becchi, Rouet and Stora \cite{BRS} and
Batalin and Vilkovisky \cite{BV}. Its essential features are the
combination of all symmetries into the BRS transformations, where the
algebraic structure is encoded in the nilpotency of the BRS operator. The
corresponding Slavnov--Taylor identity includes all symmetries  and
can be used to prove renormalizability of supersymmetric gauge
theories independent of the existence of an invariant regularization
scheme \cite{White92a,MPW96a}. The possible anomalies turn out to be
just the supersymmetric extensions of the usual gauge anomalies and
are therefore completely characterized by the gauge structure.
Furthermore, in \cite{MPW96a} it was shown that this setup leads to a
theory with the expected physical properties. One can define a set of
physical observables, i.e.~gauge invariant operators, and generators
for supersymmetry transformations and translations, and can prove that
the unmodified supersymmetry algebra is realized on the physical
observables.

In this article we consider the supersymmetric extension of QED (SQED)
as a toy model for general supersymmetric gauge theories and in
particular for the supersymmetric extensions of the standard model. 
From the Slavnov--Taylor identity we derive symmetry conditions,
simple identities between 
renormalized vertex functions. On the one hand, these conditions are
exact physical statements expressing symmetry relations, like mass
equalities and charge universality, more immediately. On the other
hand, they are used to simplify and to streamline the practical
determination of counterterms significantly. As examples we apply
these identities to various self energies and vertex corrections
calculated with DReg. We also examine  the effect of ``forgetting''  a
non-invariant but necessary counterterm. It turns out that in this
case the numerical error can significantly change the result of the
calculation.

The plan of the article is as follows: In section 2 we describe the
classical action of SQED and give its symmetries in the form of
functional identities, which are the Slavnov--Taylor identity and the
gauge Ward identity. In addition, we derive the invariant counterterms
and the corresponding normalization conditions. In section
\ref{SecSymCond} the symmetry conditions are derived. In section
\ref{SecApplications} we demonstrate in several examples, how
non-invariant counterterms appearing in DReg are identified and
removed by the use of symmetry identities. The Appendix contains the
list of the conventions used in this article. 



\section{Definition of the model}
\label{SecDefinition}

\subsection{Classical theory}

Supersymmetric QED (SQED) \cite{WZ74_SQED} is an abelian gauge theory
with the following field content: 
\begin{enumerate}
\item One vector multiplet $(A^\mu, \lambda^\alpha,
  \lambdabar^\alphadot)$ consisting of the
  photon and the photino, described by a vector and a Majorana spinor
  field. 
\item Two chiral multiplets $(\psi_L^\alpha, \phi_L)$ and 
  $(\psi_R^\alpha, \phi_R)$ with charges $Q_L = -1$, $Q_R = +1$, each
  consisting of one Weyl spinor and one scalar field, constituting
  the left- and right-handed electron and selectron, the matter
  fields.  
\end{enumerate}
The electron Dirac spinor and the photino Majorana spinor are given by
\begin{eqnarray}
\Psi = {{\psi_L}_\alpha \choose \psibar_R^\alphadot} ,\quad
\tilde{\gamma} = {-i\lambda_\alpha \choose i\lambdabar^\alphadot}\ .
\label{4Spinors}
\end{eqnarray}
The SQED Lagrangian contains kinetic, minimal coupling and mass terms
and in addition, due to the supersymmetry, coupling terms to the
photino and quartic terms in the selectron fields: 
\begin{eqnarray}
{\L}_{\rm SQED} & = &
-\frac{1}{4}F_{\mu\nu}F^{\mu\nu} +
\frac{1}{2}\overline{\tilde{\gamma}}i\gamma^\mu\partial_\mu\tilde{\gamma} 
\nonumber\\
&&{}
|D_\mu \phi_L|^2 + |D_\mu \phi_R^\dagger|^2 
+ \overline{\Psi}i\gamma^\mu D_\mu \Psi
\nonumber\\
&&{}
-\sqrt{2}eQ_L\left(
\overline{\Psi}P_R\tilde{\gamma}\phi_L
- \overline{\Psi}P_L\tilde{\gamma} \phi_R^\dagger
+ \phi_L^\dagger \overline{\tilde{\gamma}}P_L \Psi
- \phi_R \overline{\tilde{\gamma}}P_R \Psi \right)
\nonumber\\
&&{}
-\frac{1}{2}\left(eQ_L|\phi_L|^2 + eQ_R|\phi_R|^2\right)^2
\nonumber\\
&&{}
-m\overline{\Psi} \Psi - m^2 (|\phi_L|^2 + |\phi_R|^2)
\label{LSQED}
\end{eqnarray}
with the gauge covariant derivative and field strength
\begin{eqnarray}
D_\mu & = & \partial_\mu+ieQ A_\mu\ ,\\
F_{\mu\nu} & = & \partial_\mu A_\nu - \partial_\nu A_\mu\ .
\end{eqnarray}

The use of this set of physical fields corresponds to the choice of
the Wess--Zumino gauge, where unphysical fields of the vector
supermultiplet are eliminated by gauge transformations, and the
elimination of further auxiliary fields in the 
superfield version of SQED. While the former modifies the
supersymmetry algebra by gauge transformations, the second contributes
terms that vanish only if the equations of motion hold. 
In fact, the supersymmetry generators $Q_\alpha,
\overline{Q}_\alphadot$ 
satisfy 
\begin{eqnarray}
\{ Q_\alpha, \overline{Q}_\alphadot \} = 
2 P_\mu \sigma^\mu_{\alpha\alphadot} + 
\delta_\Lambda + 
\mbox{eqs.~of motion},
\label{SusyAlgebra}
\end{eqnarray}
where $\delta_\Lambda$ is an abelian gauge transformation with the
gauge function $\Lambda=-2i A_\mu\sigma^\mu_{\alpha\alphadot}$. The
equations-of-motion terms appear only when the anticommutator acts on
spinor fields.

\subsection{Quantization}

For quantizing the supersymmetric extension of QED in the Wess--Zumino
gauge one has to find symmetries which  characterize the classical
action and furthermore the
one-particle irreducible (1PI) Green functions summarized in their
generating functional $\Gamma$
\begin{eqnarray}
\Gamma = \Gamma_{\rm cl} + {\cal O}(\hbar)\ .
\end{eqnarray}
The defining symmetries of the gauge invariant action
are the abelian gauge invariance and 
$N=1$ supersymmetry. As usual, one has to add to the gauge
invariant action (\ref{LSQED}) a gauge fixing term which allows to determine a
well-defined photon propagator. The QED gauge fixing, however, breaks
the supersymmetry  non-linearly in the propagating fields and cannot be used
without modifications for a higher order construction. To overcome
this difficulty gauge and supersymmetry transformations are included into one
BRS transformation with the respective ghosts \cite{HLW90,White92a}.
It is then possible to extend the gauge fixing by a ghost part in such
a way that the complete action is  invariant under BRS transformations
(cf.~(\ref{GaugeFixingTerm}) and (\ref{GaugeFixing})).  
Moreover, by transforming also the ghosts
appropriately the algebra of supersymmetry and gauge
transformations is summarized in the nilpotency of the BRS
transformations.

For proving renormalizability it has to be shown that the Green
functions of SQED satisfy the  Slavnov--Taylor
identity, which is
the functional form of the BRS transformations, to all orders:
\begin{eqnarray}
S (\Gamma) = 0.
\end{eqnarray}
Renormalizability of N=1 supersymmetric gauge theories in the
Wess--Zumino gauge has been
proven in  \cite{MPW96a}. There and in  \cite{White92a}
it has been shown  in the framework of algebraic renormalization
that the only possible anomaly
appearing in supersymmetric gauge theories is the supersymmetric
extension of the Adler--Bardeen anomaly.
If no anomalies are present, as it is
in QED and SQED, 
all breakings  are scheme dependent breakings
and  are removed by adding appropriate counterterms. 

It is a basic fact of renormalized perturbation theory \cite{Bogoliubov}
that by the requirement of unitarity, causality and Lorentz invariance
--- leading to the usual Feynman diagram expansion ---
the higher order contributions to $\Gamma$ are not uniquely defined:
Given $\Gamma$ renormalized up to the order $\hbar^{n-1}$, the local
contributions in the next order $\hbar^n$ are ambiguous. Accordingly, 
different regularization schemes used to calculate the Feynman
diagrams can differ in the results for the local contributions, which
in general are divergent; the non-local contributions, however, are
unique and finite. That is why the ambiguity inherent in the
renormalization procedure is equivalent to the possibility to
add local counterterms of order $\hbar^n$ to $\Gamma$:
\begin{eqnarray}
\Gamma^{(n)} = \Gamma^{(n)}_{\rm{regularized}} +
\Gamma^{(n)}_{\rm{ct}} \ .
\end{eqnarray}
The divergent parts of the counterterms must cancel the divergencies
of the regularized loop diagrams whereas the finite parts are
generally only restricted by hermiticity, Lorentz invariance and power
counting renormalizability but otherwise free. 
All these counterterms may be collected and added to the classical
action:
\begin{eqnarray} 
\Gamma_{\rm eff}^{(\le n)} = \Gamma_{\rm cl} + \sum_{m=1}^n
\Gamma^{(m)}_{\rm ct} \ .
\end{eqnarray}
$\Gamma_{\rm eff}$ is the action to be used to derive the Feynman
rules of the next order $\hbar^{n+1}$, thus providing an inductive
procedure.

All conceivable finite counterterms have to be fixed by the symmetries
and by normalization conditions. Proceeding from the lowest order by
induction, all scheme-dependent breakings of the Slavnov--Taylor
identity $\Delta^{(n)}$ appearing in order $n$ have to be absorbed by
adjusting the respective non-invariant counterterms:
\begin{eqnarray}
S(\Gamma^ {(\leq n-1)}+\Gamma^{(n)}_{\rm{regularized}} +
\Gamma^{(n)}_{\rm{ct}}
 ) = \Delta^{(n)} + s_{\Gamma_{\rm cl}}
 \Gamma^{(n)}_{\rm{ct}}
= 0 + {\cal O}(\hbar^{n+1}).
\end{eqnarray}
(Here $s_{\Gamma_{\rm cl}}$ is the linearized Slavnov--Taylor operator defined
in (\ref{STLinearized}).)
At the same time this equation  fixes uniquely all non-invariant
 counterterms of  a specific scheme without referring to invariance properties
 of the scheme. 

 Since the construction of supersymmetric
 gauge theories in the Wess--Zumino gauge by means of the 
 Slavnov--Taylor identity has not been applied yet in
 phenomenological calculations, we present the construction of the symmetry
 operators and the ghost action in some detail in the following
 part of the paper.

\subsection{Symmetry requirements}
\label{SecSymReq}

The BRS formalism encodes  the complicated structure of
(\ref{SusyAlgebra}) in the simple equation
\begin{eqnarray}
s^2 = 0 + \mbox{eqs. of motion (e.o.m.)}.
\label{sSquared}
\end{eqnarray}
Here $s$ is the generator of BRS transformations given below.
In the BRS transformations the Faddeev--Popov ghost $c(x)$ is used
together with space-time independent supersymmetry and translation
ghosts $\epsilon^\alpha,\epsilonbar^\alphadot$ and $\omega^\nu$ as
parameters. The transformation rules for the ghosts themselves are
given by the structure constants of the symmetry algebra
\cite{BRS}. That yields the following explicit form of the operator $s$:
\begin{eqnarray}
sA_\mu & = & \partial_\mu c + i\epsilon\sigma_\mu\lambdabar
             -i \lambda\sigma_\mu\epsilonbar -i\omega^\nu\partial_\nu A_\mu
\ ,\\
s\lambda^\alpha & = & \frac{i}{2} (\epsilon\sigma^{\rho\sigma})^\alpha
             F_{\rho\sigma} - i\epsilon^\alpha\,
             eQ_L(|\phi_L|^2-|\phi_R|^2) -i\omega^\nu\partial_\nu  
             \lambda^\alpha
\ ,\\
s\lambdabar_\alphadot & = & \frac{-i}{2} (\epsilonbar\sigmabar^{\rho\sigma})
             _\alphadot F_{\rho\sigma} - i\epsilonbar_\alphadot\, 
             eQ_L(|\phi_L|^2-|\phi_R|^2) 
             -i\omega^\nu\partial_\nu \lambdabar_\alphadot 
\ ,\\
s\phi_L & = & -ieQ_L c\,\phi_L +\sqrt2\, \epsilon\psi_L - i\omega^\nu\partial_\nu \phi_L
\ ,\\
s\phi_L^\dagger & = & +ieQ_L c\,\phi_L^\dagger +\sqrt2\, \psibar_L\epsilonbar - i\omega^\nu\partial_\nu \phi_L^\dagger
\ ,\\
s\psi_L^\alpha & = & -ieQ_L c\,\psi_L^\alpha - \sqrt2\, 
         \epsilon^\alpha\, m\phi_R^\dagger 
         -\sqrt2\, i (\epsilonbar\sigmabar^\mu)^\alpha D_\mu\phi_L 
         -i\omega^\nu\partial_\nu \psi_L^\alpha
\ ,\\
s{\psibar_L}_\alphadot & = & +ieQ_L c\,{\psibar_L}_\alphadot 
         + \sqrt2\,\epsilonbar_\alphadot\, m\phi_R + \sqrt2\, i
         (\epsilon\sigma^\mu)_\alphadot (D_\mu\phi_L)^\dagger 
         -i\omega^\nu\partial_\nu {\psibar_L}_\alphadot
\ ,\\
sc & = & 2i\epsilon\sigma^\nu\epsilonbar A_\nu -i\omega^\nu\partial_\nu c
\ ,\\
s\epsilon^\alpha & = & 0
\ ,\\
s\epsilonbar^\alphadot & = &0
\ ,\\
s\omega^\nu & = & 2\epsilon\sigma^\nu\epsilonbar
\ ,\\
s\bar c & = & B - i\omega^\nu\partial_\nu \bar c
\ ,\\
sB & = & 2i\epsilon\sigma^\nu\epsilonbar \partial_\nu \bar c 
         -i\omega^\nu\partial_\nu B
\end{eqnarray}
and corresponding transformations for the right-handed fields. Here we
have introduced also the antighost $\bar{c}$ and the auxiliary field
$B$ appearing in the gauge fixing in later course (see
eq.~(\ref{GaugeFixingTerm})).

The symmetries of the classical Lagrangian are summarized in the
equation 
\begin{eqnarray}
s\Gamma_{\rm SQED} = 0 
\label{Invariance}
\end{eqnarray}
for $\Gamma_{\rm SQED} = \intx \L_{\rm SQED}$.

The remaining obstructions are the non-linear BRS transformations and
the eqs.-of-motion terms in the nilpotency of $s$. Both are overcome
by using external fields. Each non-linear BRS transformation
$s\varphi_i$ is coupled to an external field $Y_i$:
\begin{eqnarray}
\Gamma_{\rm ext} & = & \intx \Bigl(Y_\lambda^\alpha s\lambda_\alpha
+ Y_\lambdabar{}_\alphadot s\lambdabar^\alphadot
\nonumber\\&&{}\hfill
+ Y_{\phi_L} s\phi_L + Y_{\phi_L^\dagger} s\phi_L^\dagger
+ Y_{\psi_L}^\alpha s\psi_L{}_\alpha
+ Y_{\psibar_L}{}_\alphadot s\psibar_L^\alphadot
+ ({}_{L\to R})\Bigr) \ .
\end{eqnarray}
The statistics, dimension and ghost number of the $Y_i$ is such that
$\Gamma_{\rm ext}$ has the same quantum numbers as $\Gamma_{\rm
  SQED}$. In this way we can use the $Y_i$ as sources for the
non-linear BRS transformations and write $s\varphi_i =
\delta\Gamma_{\rm ext}/\delta Y_i$, where the r.h.s.~possesses a
well-defined extension to higher orders. Moreover, as was 
realized in \cite{BV}, it is possible to extend
the classical action by terms bilinear in the sources that absorb the
eqs.-of-motion terms. Hence, the sum 
\begin{eqnarray}
\Gamma_{\rm cl} & = &\Gamma_{\rm SQED} + \Gamma_{\rm ext} + \Gamma_{\rm
  bil} \ ,\\
\Gamma_{\rm bil} & = & -(Y_\lambda \epsilon)(\epsilonbar Y_\lambdabar)
                   -2(Y_{\psi_L} \epsilon)(\epsilonbar  Y_{\psibar_L})
                   -2(Y_{\psi_R} \epsilon)(\epsilonbar  Y_{\psibar_R})
\end{eqnarray}
satisfies the Slavnov--Taylor identity
\begin{eqnarray}
S(\Gamma_{\rm cl}) & = & 0
\ .
\end{eqnarray}
The Slavnov--Taylor operator acting on a general functional ${\cal F}$
is defined as
\begin{eqnarray}
S({\cal F}) & = & 
\intx\Bigl(s A^\mu \dF{A^\mu} + s c \dF{c} + s\bar{c} \dF{\bar{c}}
           + s B \dF{B}
\nonumber\\&&{}\quad
+ \dF{Y_\lambda{}_\alpha}\dF{\lambda^\alpha}
+ \dF{Y_\lambdabar^\alphadot}\dF{\lambdabar_\alphadot}
\nonumber\\&&{}\quad
+ \dF{Y_{\phi_L}}\dF{\phi_L}
+ \dF{Y_{\phi_L^\dagger}}\dF{\phi_L^\dagger}
+ \dF{Y_{\psi_L{}_\alpha}}\dF{\psi_L^\alpha}
+ \dF{Y_{\psibar_L}^\alphadot}\dF{\psibar_L{}_\alphadot}
+(_{L\to R})\Bigr)
\nonumber\\&&{}
+ s\epsilon^\alpha\frac{\partial{\cal F}}{\partial\epsilon^\alpha}
+ s\epsilonbar_\alphadot
  \frac{\partial{\cal F}}{\partial\epsilonbar_\alphadot}
+ s\omega^\nu \frac{\partial{\cal F}}{\partial\omega^\nu}
\label{STOperator}
\nonumber
\\& \equiv &
\int \Bigl(s\varphi_i' \frac{\delta{\cal F}}{\delta \varphi_i'}
+  \frac{\delta{\cal F}}{\delta Y_i} 
\frac {\delta{\cal F}}{\delta \varphi_i}\Bigr) \ .
\end{eqnarray}
In the last line a symbolic abbreviation has been introduced in which
$\varphi_i'$ runs over all linearly transforming fields and the global
ghosts. The electron contributions to $\Gamma_{\rm ext}$ and $S({\cal
  F})$ can be written in terms of 4-spinors as 
\begin{eqnarray}
\Gamma_{\rm ext}|_\Psi & = & \intx \left(Y_{\Psi} s\Psi +
Y_\Psibar s\Psibar \right)
\ ,\\
S({\cal F})|_\Psi & = & \intx \Bigl(\dF{Y_\Psi}\dF{\Psi} + 
\dF{Y_{\Psibar}}\dF{\Psibar}\Bigr)
\end{eqnarray}
with the 4-spinors from eq.~(\ref{4Spinors}) and
\begin{eqnarray}
Y_\Psi & = & \left(Y_{\psi_L}{}^\alpha , Y_{\psibar_R}{}_\alphadot
\right)
\ ,\quad
Y_\Psibar = {-Y_{\psi_R}{}_\alpha \choose -Y_{\psibar_L}{}^\alphadot}
\ ,\\
\dfunc{}{Y_\Psi} & = & {\dfunc{}{Y_{\psi_L}{}^\alpha} \choose 
                      \dfunc{}{Y_{\psibar_R}{}_\alphadot}}
\ ,\quad
\dfunc{}{Y_{\Psibar}} = \left(\dfunc{}{Y_{\psi_R}{}_\alpha},
                            \dfunc{}{Y_{\psibar_L}{}^\alphadot}\right)
\ .
\end{eqnarray}

The Slavnov--Taylor identity is the key for solving the above
mentioned problems since it may be extended to higher orders and it
contains both the invariance (\ref{Invariance}) and the nilpotency
(\ref{sSquared}): The invariance of $\Gamma_{\rm SQED}$ is expressed
in the terms without $Y_j$, and the terms linear in the $Y_j$ express
the symmetry algebra acting on the corresponding fields $\varphi_j$: 
\begin{eqnarray}
(Y_j)^0:&\makebox[8.2cm][l]{$\displaystyle\ 
\int\Bigl( s\varphi_i' \frac{\delta\Gamma_{\rm SQED}}{\delta \varphi_i'}
+ \frac{\delta\Gamma_{\rm ext}}{\delta Y_i} 
\frac {\delta\Gamma_{\rm SQED}}{\delta \varphi_i}\Bigr) \hfill\to$} &
\  s\Gamma_{\rm SQED} = 0 \ ,\\
(Y_j):&\makebox[8.2cm][l]{$\displaystyle\ 
\int\Bigl( s\varphi_i' \frac{\delta\Gamma_{\rm ext}}{\delta \varphi_i'}
+ \frac{\delta\Gamma_{\rm ext}}{\delta Y_i} 
\frac {\delta\Gamma_{\rm ext}}{\delta \varphi_i} +
\frac{\delta\Gamma_{\rm bil}}{\delta Y_i} 
\frac {\delta\Gamma_{\rm SQED}}{\delta \varphi_i}\Bigr)  \hfill\to$} &
\ s^2 \varphi_j=\mbox{ e.o.m.}
\end{eqnarray}
The linearized Slavnov--Taylor operator, defined for bosonic
functionals ${\cal F}$, is given by
\begin{eqnarray}
s_{\cal F} & = & \int\Bigl( s\varphi_i' \frac{\delta}{\delta \varphi_i'}
+  \frac{\delta{\cal F}}{\delta Y_i} 
\frac {\delta}{\delta \varphi_i}   
+  \frac {\delta{\cal F}}{\delta \varphi_i}   
\frac{\delta}{\delta Y_i} \Bigr)
\ .
\label{STLinearized}
\end{eqnarray}
The full Slavnov--Taylor operator and its linearized version have the
nilpotency property 
\begin{eqnarray}
s_{\cal F} S({\cal F}) & = & 0
\label{Jacobi2}
\end{eqnarray}
if the functional ${\cal F}$ satisfies the linear identity
\begin{eqnarray}
i\epsilon\sigma^\mu\frac{\delta\cal F}{\delta  Y_\lambdabar}
-i\frac{\delta\cal F}{\delta Y_\lambda}\sigma^\mu\epsilonbar
+i\omega^\nu\partial_\nu(i\epsilon\sigma^\mu\lambdabar
                        -i\lambda\sigma^\mu\epsilonbar)
-2i\epsilon\sigma_\nu\epsilonbar F^{\nu\mu} = 0\ ,
\label{EquivNilpotency}
\end{eqnarray}
which is equivalent to nilpotency on $A^\mu$: $s_{\cal F}^2 A^\mu=0$. 
Eq.~(\ref{EquivNilpotency}) is satisfied in particular by $\Gamma_{\rm
  cl}$. 

The gauge fixing term has to be chosen in such a way that
renormalizability by power-counting is ensured. We define
\begin{eqnarray}
\Gamma_{\rm fix} & = & \intx\ s_{\Gamma_{\rm cl}}
          [\bar{c}( \partial^\mu A_\mu + \frac{\xi}{2} B)] 
\nonumber\\
& = &
 \intx\Bigl( B\partial^\mu A_\mu + \frac{\xi}{2}B^2
- \bar{c}\Box c
\nonumber\\
&&{}\quad
- \bar{c}\partial^\mu(i\epsilon\sigma_\mu\lambdabar 
                     -i\lambda\sigma_\mu\epsilonbar) 
+ \xi i \epsilon\sigma^\nu\epsilonbar
  (\partial_\nu\bar{c})\bar{c} \Bigr)
\label{GaugeFixingTerm}
\end{eqnarray}
with a real gauge parameter $\xi$. This gauge fixing term is added to
the classical action: 
\begin{eqnarray}
\Gamma_{\rm cl} \to \Gamma_{\rm cl} + \Gamma_{\rm fix}\ .
\label{GaugeFixing}
\end{eqnarray}
Introducing the gauge fixing in this way the
Slavnov--Taylor identity remains valid. Indeed we see
that in addition to the usual QED gauge fixing and  ghost terms, which
break supersymmetry, there arise compensating terms dependent
on the constant ghost fields $\epsilon,\epsilonbar$.

\paragraph{Symmetry requirements on $\Gamma$:}
The symmetry properties of $\Gamma_{\rm cl}$ are now imposed as
constraints on $\Gamma$. In addition to the Slavnov--Taylor identity
several linear equations and manifest symmetries are imposed. To
summarize: 
\begin{itemize}
\item Slavnov--Taylor identity and nilpotency of $s_\Gamma$:
\begin{eqnarray}
S(\Gamma) & = & 0\ ,
\label{STI}
\\
s_\Gamma^2 A^\mu & = & 0\ .
\label{Nilpotency}
\end{eqnarray}
The latter condition is equivalent to eq.~(\ref{EquivNilpotency}) for
${\cal F}=\Gamma$, and according to eq.~(\ref{Jacobi2}) it is already
sufficient for the nilpotency relation $s_\Gamma S(\Gamma) = 0$. 
\item Gauge fixing condition, ghost equations:
\begin{eqnarray}
&&
\dg{B} = \frac{\delta\Gamma_{\rm cl}}{\delta B},\quad
\dg{c} = \frac{\delta\Gamma_{\rm cl}}{\delta c},\quad
\dg{\omega^\mu} = \frac{\delta\Gamma_{\rm cl}}{\delta \omega^\mu},\quad
\dg{\bar{c}} = \frac{\delta\Gamma_{\rm cl}}{\delta \bar{c}} \ .
\label{GEQ}
\end{eqnarray}
It is possible to require that these derivatives do not receive quantum
corrections since they are linear in the dynamical fields at the
classical level. These equations serve as normalization
conditions; their physical consequences are explained in the next
subsection. 
\item Manifest symmetries: We require $\Gamma$ to be invariant under
  the discrete symmetries $R, C, CP$ and to be electrically and ghost
  charge neutral, Lorentz invariant and bosonic. The quantum numbers
  of the fields are determined by the
  corresponding symmetries of $\Gamma_{\rm cl}$ and are listed in
  tab.~\ref{TabDiscreteSym}, \ref{TabQNumbers}. 
  Note that the usual $R$-parity is the same as our $R^2$ and thus
  less restrictive than our $R$. Contrary to the preceding
  symmetries, we assume these ones to be manifestly preserved, which is
  true for all common regularization schemes.
\begin{table}[tbh]
\begin{displaymath}
\begin{array}{|c||c|c|c|c|c|c|c|c|c|c|c|c|c|c|c|c|c|c|}
\hline
\chi & x^\mu & A^\mu  & -i\lambda^\alpha &  
\phi_{L} & \phi_{R} &
\psi^\alpha_{L} &  
\psi^\alpha_{R} &   c & \epsilon^\alpha & 
\omega^\nu & \bar{c} & B 
\\ \hline
R\chi & x^\mu & A^\mu & -\lambda^\alpha & 
-i\phi_L & -i\phi_R & 
\psi^\alpha_{L} & 
\psi^\alpha_{R} &   c & -i\epsilon^\alpha & 
\omega^\nu & \bar{c} & B 
\\ \hline
C\chi & x^\mu & -A^\mu  & i\lambda^\alpha &   
\phi_{R} & \phi_{L} &
\psi^\alpha_{R} &  
\psi^\alpha_{L} &  -c & \epsilon^\alpha & 
\omega^\nu & -\bar{c} & -B 
\\ \hline
CP\chi & ({\cal P}x)^\mu & -({\cal P}A)^\mu  & -\lambdabar_\alphadot &
\phi_{L}^\dagger &\phi_{R}^\dagger &
i\psibar_L{}_\alphadot &  
i\psibar_R{}_\alphadot &   -c & -i\epsilonbar_\alphadot &
 ({\cal P}\omega)^\nu & -\bar{c} & -B 
\\ \hline
\end{array}
\end{displaymath}
{\caption{{Discrete symmetries.}
The transformation rules for the sources $Y_i$ can be deduced from the
requirement that $\Gamma_{\rm ext}$ is invariant and the
transformation rules for the complex conjugate fields are obvious
except for the $CP$ conjugation of the spinors. We define 
for $\chi\in\{\lambda,\psi_L,\psi_R,\epsilon\}:$
\hspace*{\fill}
\mbox{\hspace*{0cm}}
\hspace*{\fill}
\mbox{$\chi^\alpha \stackrel{CP}{\to} a \chibar_\alphadot\Rightarrow
\chibar_\alphadot \stackrel{CP}{\to} -a^* \chi^\alpha \ ,\ 
\chi_\alpha \stackrel{CP}{\to} -a \chibar^\alphadot\ , \ 
\chibar^\alphadot \stackrel{CP}{\to} a^* \chi_\alpha 
\ .$}
\hspace*{\fill}
\label{TabDiscreteSym}
}}
\end{table}
\begin{table}[tbh]
\begin{displaymath}
\begin{array}{|c||c|c|c|c|c|c|c|c|c|c|c|c|c|c|c|c|c|c|}
\hline
\chi & x^\mu & A^\mu  & -i\lambda^\alpha &  
\phi_{L} & \phi_{R} &
\psi^\alpha_{L} &  
\psi^\alpha_{R} &   c & \epsilon^\alpha & 
\omega^\nu & \bar{c} & B 
\\ \hline
Q   & 0  &  0  & 0 & -1 & +1 & -1 & +1 & 0 & 0 & 0 & 0 & 0 \\ \hline 
Q_c & 0  &  0  & 0 &  0 & 0  &  0 &  0 &+1 & +1&+1 &-1 & 0 \\ \hline
GP  & 0  &  0  & 1 &  0 & 0  &  1 &  1 & 1 & 0 & 1 & 1 & 0 \\ \hline
dim & -1 &  1  &3/2&  1 & 1  & 3/2& 3/2& 0 &-1/2& -1& 2 & 2 \\ \hline
\end{array}
\end{displaymath}
\caption{Quantum numbers. $Q,Q_c,GP,dim$ denote electrical and ghost
  charge, Grassmann parity and the mass dimension, respectively. The
  quantum numbers of the sources $Y_i$ can be obtained from the
  requirement that $\Gamma_{\rm ext}$ is neutral, bosonic and has
  $dim=4$. The commutation rule for two general fields is
  $\chi_1\chi_2 = (-1)^{GP_1 GP_2} \chi_2\chi_1$.}
\label{TabQNumbers} 
\end{table}
\end{itemize}

\subsection{Immediate consequences}

The conditions for $\bar{c}$ and $B$ in eq.~(\ref{GEQ}) forbid any
quantum corrections to $\Gamma_{\rm fix}$ and thus play the role of
gauge fixing conditions. The ghost equations in eq.~(\ref{GEQ}) for
$c,\omega^\mu$ have a direct physical consequence: They imply, in
connection with the Slavnov--Taylor identity, Ward identities for
electrical current conservation and translational invariance. This can
be seen from the following consistency equations for general bosonic
functionals ${\cal F}$: 
\begin{eqnarray}
\dfunc{}{c} S({\cal F}) +s_{\cal F}\frac{\delta\cal F}{\delta c} 
& = & 
-\partial^\mu\frac{\delta \cal F}{\delta A^\mu} 
- i\omega^\nu\partial_\nu\frac{\delta \cal F}{\delta c}  
\ ,\\
\dfunc{}{\omega^\mu} S({\cal F}) 
+s_{\cal F}\frac{\delta \cal  F}{\delta \omega^\mu}  
& = & 
-i\int(\partial_\mu\varphi_i') \frac{\delta \cal F}{\delta \varphi_i'}
\ ,\\
\dfunc{}{\bar{c}} S({\cal F}) 
+s_{\cal F}\frac{\delta\cal F}{\delta \bar{c}} 
& = & 
-2i\epsilon\sigma^\nu\epsilonbar \partial_\nu \frac{\delta\cal
  F}{\delta B}
-i\omega^\nu\partial_\nu \frac{\delta \cal F}{\delta \bar{c}}
\ ,\\
\dfunc{}{B} S({\cal F}) - s_{\cal F}\frac{\delta\cal F}{\delta B}
& = & 
 \frac{\delta \cal F}{\delta \bar{c}} 
+ i\omega^\nu\partial_\nu \frac{\delta\cal F}{\delta B}
\ .
\end{eqnarray}
For ${\cal F} = \Gamma$ and $S(\Gamma)=0$ the first two equations lead
to the announced Ward identities:
\begin{eqnarray}
\partial^\mu\dg{A^\mu} &  = & -ie w_{\rm em}\Gamma -\Box B
 +{\cal O}(\omega)
\label{WI}
\ ,\\
w_{\rm em} & = & Q_L\Bigl(\phi_L\dfunc{}{\phi_L} -
Y_{\phi_L}\dfunc{}{Y_{\phi_L}} + \psi_L\dfunc{}{\psi_L}
- Y_{\psi_L}\dfunc{}{Y_{\psi_L}} 
\nonumber\\&&{}
- \phi_L^\dagger\dfunc{}{\phi_L^\dagger}
 + Y_{\phi_L^\dagger}\dfunc{}{Y_{\phi_L^\dagger}}
 - \psibar_L\dfunc{}{\psibar_L}
 - Y_{\psibar_L}\dfunc{}{Y_{\psibar_L}} \Bigr) 
\nonumber\\&&{}+ (_{L\to R})
\end{eqnarray}
and
\begin{eqnarray}
0 & = & \intx\Bigl( \partial_\mu\varphi_i' \dg{\varphi_i'}
           + \partial_\mu\varphi_i \dg{\varphi_i}
           + \partial_\mu Y_i \dg{Y_i}\Bigr)
\ .
\end{eqnarray}
The $\omega$-dependent terms in the electromagnetic Ward identity
(\ref{WI}) arise because translations do not commute with local gauge
transformations.

Conversely, if the Ward identities and the linear
eqs.~(\ref{Nilpotency}), (\ref{GEQ}) hold, the consistency equations
yield 
\begin{eqnarray}
\dfunc{}{c} S(\Gamma) & = & \dfunc{}{\omega^\nu} S(\Gamma) =
\dfunc{}{\bar{c}} S(\Gamma) =
\dfunc{}{B} S(\Gamma)  =  0\ .
\end{eqnarray}
In this case, therefore, the Slavnov--Taylor identity can not be
broken by terms depending on $c,\omega^\nu,\bar{c},B$. 

\subsection{Most general symmetric counterterms}
\label{SectionSymmetricCTs}

The symmetry requirements fix $\Gamma$ up to additive symmetric
counterterms in each order. To find the symmetric counterterms
we take two solutions $\Gamma$ and $\tilde{\Gamma} = \Gamma
+ \zeta \Gamma_{\rm sym}$ of the symmetry requirements at first order
in the infinitesimal parameter $\zeta$ and calculate the most general
counterterms $\Gamma_{\rm sym}$. The requirements that the
Slavnov--Taylor identity eq.~(\ref{STI}) is satisfied at first order
in $\zeta$ can be cast into the form
\begin{eqnarray}
s_{\Gamma_{\rm cl}} \Gamma_{\rm sym} = 
\int\Bigl( s\varphi_i'\frac{\delta\Gamma_{\rm sym}}{\delta\varphi_i'}
 + \frac{\delta\Gamma_{\rm cl}}{\delta Y_i} 
\frac {\delta\Gamma_{\rm sym}}{\delta \varphi_i} 
+
\frac{\delta\Gamma_{\rm sym}}{\delta Y_i} 
\frac {\delta\Gamma_{\rm cl}}{\delta \varphi_i}\Bigr)  =  0\ ,
\end{eqnarray}
and eq.~(\ref{GEQ}) prevents a dependence of
$\Gamma_{\rm sym}$ on $B,c,\omega^\nu,\bar{c}$.
The solution reads
\begin{eqnarray}
\Gamma_{\rm sym} & = & 
\Bigl[\delta Z_m m\frac{\partial}{\partial m} 
\nonumber\\&&{}
+ \frac{1}{2}\delta Z_\gamma \Bigl(- e\frac{\partial}{\partial e}
+ 2\xi\frac{\partial}{\partial\xi} 
\nonumber\\&&{}\quad
+ \intx\Bigl(A^\mu\dfunc{}{A^\mu} 
+ \lambda^\alpha\dfunc{}{\lambda^\alpha}
 - Y_{\lambda}{}^\alpha\dfunc{}{Y_{\lambda}{}^\alpha}
+ \lambdabar_\alphadot\dfunc{}{\lambdabar_\alphadot}
 - Y_{\lambdabar}{}_\alphadot\dfunc{}{Y_{\lambdabar}{}_\alphadot}
\nonumber\\&&{}\quad
+ c\dfunc{}{c} 
- \bar{c}\dfunc{}{\bar c} 
- B\dfunc{}{B} \Bigr)\Bigr)
\nonumber\\&&{}
+ \frac{1}{2}\delta Z_{\phi}\intx\Bigl(\phi_L\dfunc{}{\phi_L}
 - Y_{\phi_L}\dfunc{}{Y_{\phi_L}}
+  \phi_R\dfunc{}{\phi_R}
 - Y_{\phi_R}\dfunc{}{Y_{\phi_R}} 
\nonumber\\&&{}\quad
+ \phi^\dagger_L\dfunc{}{\phi^\dagger_L}
 - Y_{\phi^\dagger_L}\dfunc{}{Y_{\phi^\dagger_L}}
+ \phi^\dagger_R\dfunc{}{\phi^\dagger_R}
 - Y_{\phi^\dagger_R}\dfunc{}{Y_{\phi^\dagger_R}} \Bigr)
\nonumber\\&&{}
+ \frac{1}{2}\delta Z_\Psi\intx \Bigl(\Psi\dfunc{}{\Psi}
 - Y_{\Psi}\dfunc{}{Y_{\Psi}}
+ \Psibar\dfunc{}{\Psibar}
 - Y_{\Psibar}\dfunc{}{Y_{\Psibar}} \Bigl)
\Bigr]  \    \Gamma_{\rm cl} 
\label{SymCT}
\end{eqnarray}
with four free constants $\delta Z_m$, $\delta Z_\gamma$, 
$\delta Z_\phi$, $\delta Z_\Psi$. The 
condition for $\dg{c}$ in (\ref{GEQ}) and the Ward identity
(\ref{WI}) result in $e$ being the effective charge in the Thomson
limit (see section \ref{SecPhysCond}) and thus prevent an independent
charge renormalization. The action of this differential
operator on the classical action just corresponds to a
multiplicative renormalization of the parameters and fields appearing
therein. That means that after restoring the
symmetries all divergencies from the loop diagrams may be absorbed by
redefinitions of the parameters and fields appearing in $\Gamma_{\rm
  cl}$, which is the usual understanding of multiplicative
renormalizability.  

\subsection{Normalization conditions}

To fix the remaining ambiguity of the symmetric counterterms we impose
the usual normalization conditions\footnote{In the literature also
  labeled as ``renormalization conditions''.} for QED. These are 
on-shell normalization conditions for the mass parameter and the
photon self energy, and conditions at an arbitrary scale $\kappa$ for
the normalization of the matter self energies:\footnote{
$\kappa^2=m^2$ would lead to infrared divergences in the normalization
conditions.}
\begin{eqnarray}
\Gamma_{\phi_L\phi_L^\dagger}(-p,p) & = & 0 \mbox{ for } p^2=m^2
\ ,\\
\lim_{p^2\to0}\frac{1}{p^2}\Gamma_{A^\mu
  A^\nu}(-p,p)|_{g_{\mu\nu}-\rm part} & = & -g_{\mu\nu}
\label{NormPhoton}
\ ,\\
\frac{\partial}{\partial p^2} \Gamma_{\phi_L\phi_L^\dagger}(-p,p) 
 & = & 1 \mbox{ for } p^2=\kappa^2
\ ,\\
\Gamma_V(p^2) + 2m^2(\Gamma_V^\prime(p^2) -\Gamma_S^\prime(p^2)) & = &
1  \mbox{ for } p^2=\kappa^2
\ . 
\end{eqnarray}
Here we have used a covariant decomposition for the electron self
energy: 
\begin{eqnarray}
\Gamma_{\Psi\Psibar}(p,-p) & = &
\psl \Gamma_V(p^2)
-
m \Gamma_S(p^2)
\end{eqnarray}
with scalar functions $\Gamma_{V,S}$. 
Since these normalization conditions have a unique classical
(i.e.~tree level) solution, they fix $\Gamma$ uniquely to all orders.

We did not require the residua of the matter propagators to be
unity on-shell. It is useful to define the functions
\begin{eqnarray}
\label{ZFactorsStart}
Z_\phi(p^2) & = & \left(\partial_{p^2} \Gamma_{\phi_L\phi_L^\dagger}(-p,p)
\right)^{-1} \ ,\\
Z_\Psi(p^2) & = & \left(\Gamma_V(p^2) + 2m^2(\Gamma_V^\prime(p^2) -
\Gamma_S^\prime(p^2)) \right)^{-1} \ .
\label{ZFactorsEnd}
\end{eqnarray}
They approach the usual (infrared divergent) $Z$ factors in the limit
$p^2\to m^2$ and appear in the LSZ reduction formula as wave function
renormalization factors:
\begin{eqnarray}
S_{fi} = \lim_{p^2\to m^2} \left(i{Z_\phi}^{-1/2}(p^2)
(-p^2+m^2) \ldots
\langle0|T\phi\ldots|0\rangle(p,\ldots)\right) \ .
\end{eqnarray}
For the present paper they play a role in the symmetry conditions
derived in the next section.



\section{Symmetry conditions}
\label{SecSymCond}

The Slavnov--Taylor identity (\ref{STI}) is a complicated non-linear
equation for the effective action with an enormous information
content. In this section we will show that it is possible to obtain
much simpler symmetry conditions as a consequence of the
Slavnov--Taylor identity and the normalization conditions. One virtue
of these symmetry conditions is that they are well suited for
practical applications. Together with the normalization conditions and
the conditions in eqs.~(\ref{Nilpotency}), (\ref{GEQ}) they form a
complete set of simple identities that determine the counterterms of
all power-counting renormalizable interactions. A similar strategy was
applied by \cite{Grassi98} in the context of the abelian Higgs-Kibble
model.

We begin this section with a particularly simple symmetry condition, to
illustrate our general method. This example also shows that is
useful to divide the symmetry conditions into two parts: the ones for
vertex functions containing external sources, expressing the higher
order modifications to the symmetry transformations, and the ones for
the vertex functions for physical fields.

Let us make some remarks on our notation and conventions. 
The manifest symmetries are always implicitly used, in particular
R-parity violating vertex functions are not mentioned and the
conditions involving one selectron field are only given for one of the
fields $\phi_L$, $\phi_R$, $\phi_L^\dagger$, $\phi_R^\dagger$. Since
it is easier to work with fields of a definite R-parity the 2-spinors
$\lambda,\lambdabar,\epsilon,\epsilonbar$ and the 4-spinors
$\Psi,\Psibar$ are used during the derivations and only for the final
results either a pure 2-spinor or a pure 4-spinor notation is
chosen. Most of the following identities stem from some derivative 
of the Slavnov--Taylor identity $\delta
S(\Gamma)/\delta\chi_1\ldots\delta\chi_n=0$, leading to products of
the form 
\begin{eqnarray}
\Gamma_{\chi_{1}\ldots\chi_{m} Y_i}(p_1,\ldots,p_m,-p)
\Gamma_{\chi_{m+1}\ldots\chi_{n} \varphi_i}(p_{m+1},\ldots,p_n,p)
\end{eqnarray}
with 
$p = p_1 + \ldots + p_m = - p_{m+1} - \ldots - p_n$
due to momentum conservation. The definition of the vertex functions
is given in app.~\ref{AppVertexFcts}. Because this structure is
general, the momenta in the arguments are not always written down
explicitly. 

\subsection{Electron--selectron mass identity}
\label{SecMass}

The normalization condition
\begin{eqnarray}
\Gamma_{\phi_L\phi_L^\dagger}(-p,p)  =  0 \mbox{ for } p^2=m^2
\label{NormPhiPhi}
\end{eqnarray}
defines $m$ to be the physical selectron mass. Using the
Slavnov--Taylor identity we will now prove the following symmetry
condition: 
\begin{eqnarray}
\Gamma_{\Psi\Psibar}(p,-p) u(p)  =  0 \mbox{ for } p^2=m^2\ ,
\label{SymPsiPsi}
\end{eqnarray}
where $u(p)$ is a spinor satisfying the Dirac equation
$(\psl-m)u(p)=0$. Physically this means that $m$ is equal to the
physical electron mass, and thus the electron and selectron masses are
equal.

The strategy for the proofs of the symmetry
conditions is first to obtain identities between vertex functions in
the usual way taking suitable derivatives of the Slavnov--Taylor
identity and setting all fields to zero afterwards. These non-linear
identities can then be solved for particular vertex functions and
further simplified if one evaluates them at the special momenta of the
normalization conditions.

Since the condition we want to prove is due to supersymmetry, we use
one derivative with respect to $\epsilon$:
\begin{eqnarray}
\dpartial{}{\epsilon}\dfunc{2}{\phi_L^\dagger(-p)\delta\Psi(p)}
S(\Gamma) |_{\varphi_i=Y_i=0}
= 0 \ .
\end{eqnarray}
After setting all fields to zero most of the terms vanish due to charge
non-conservation, and only two terms contribute:
\begin{eqnarray}
\Gamma_{\Psi\epsilon Y_{\phi_L}}(p,-p)
\Gamma_{\phi_L^\dagger\phi_L}(-p,p)
 + \Gamma_{\phi_L^\dagger\epsilon Y_{\Psibar}}(-p,p)
\Gamma_{\Psi\Psibar}(p,-p)
 = 0 \ .
\label{MassIdentity}
\end{eqnarray}
For $p^2=m^2$ the normalization condition (\ref{NormPhiPhi}) and
$\Gamma_{\phi_L^\dagger\epsilon Y_{\Psibar}}\ne0$ show that the spinor
matrix $\Gamma_{\Psi\Psibar}(p,-p)$ has the eigenvalue zero and thus
cannot be invertible. Since it must be built out of the covariants $1$
and $\psl$ it can only be proportional to $(\psl-m)$ or
$(\psl+m)$. Taking into account the lowest order result the second
possibility is excluded and the announced result (\ref{SymPsiPsi})
follows.

\subsection{Higher order supersymmetry}
\label{SecHigherOrderSusy}

Eq.~(\ref{MassIdentity}) exhibits a general feature of the
equations derived below, namely the appearance of prefactors that are
themselves vertex functions with external sources and ghost
fields, reflecting the non-linearity of the
Slavnov--Taylor identity. Their physical meaning is to represent
renormalized higher order corrections to the symmetry transformations
coupled to the sources in $\Gamma_{\rm ext}$, which will be
explained in more detail in sec.~\ref{SecSusy1Loop}. It is necessary
to derive symmetry conditions for such vertex functions before we are
able to derive further identities for vertex functions involving only
physical fields.

In fact, all vertex functions involving external $c$ or
$\omega^\mu$ ghosts --- expressing the exact gauge
transformations and translations --- are already fixed to all orders
by the requirements in eq.~(\ref{GEQ}). The vertex functions involving
external $\epsilon$ ghosts and $Y$ fields express the supersymmetry
transformations. They may acquire higher order corrections, but it is
still  possible to derive symmetry conditions constraining these
modifications because the symmetry algebra is fixed to all orders.

First we derive the supersymmetry transformations of the photino,
i.e.~the vertex functions with external $\epsilon$ and
$Y_{\lambda}$. There are only three terms of dimension $\le4$
possible: $Y_\lambda\epsilon A^\mu$, $Y_\lambda\epsilon
|\phi_{L,R}|^2$, $Y_\lambda\epsilon Y_\lambdabar\epsilonbar$ and their
CP-conjugates. To constrain the first one we use the nilpotency on
$A^\mu$, which expresses the supersymmetry algebra:
\begin{eqnarray}
0 & = & \dpartial{2}{\epsilonbar \partial\epsilon}\dfunc{}{A^\rho} 
s_\Gamma^2 A^\mu
\\
\Rightarrow
0 & = & i\Gamma_{A^\rho \epsilon^\beta Y_\lambda{}_\alpha}
        \sigma^\mu_{\alpha\betadot}
+       i\sigma^\mu_{\beta\alphadot}
        \Gamma_{A^\rho\epsilonbar^\betadot Y_\lambdabar{}_\alphadot}
\nonumber\\&&{}
+  2\psl_{\beta\betadot} g_\rho{}^\mu - 2p^\mu
\sigma_\rho{}_{\beta\betadot} 
\ .
\end{eqnarray}
The first line contains products of the transformation of the
photon into a photino and vice versa, the second line a sum of a translation
and a gauge transformation. The explicit $\sigma$ matrices originate from
$\dpartial{2}{\epsilon\partial\lambdabar}sA^\mu$ and $h.c.$, and the
terms in the second line from the BRS transformations of the $\omega$
and $c$ ghosts in $sA^\mu$. All terms are fixed except for the ones
containing $Y_\lambda, Y_\lambdabar$. Taking into account the Ward
identity (\ref{WI}), leading to $p^\rho \Gamma_{A^\rho \epsilon
  Y_\lambda}=0$, implies in connection with $CP$ invariance:
\begin{eqnarray}
\Gamma_{A^\mu \epsilon^\beta Y_\lambda{}_\alpha}(p,-p) = 
p^\rho (\sigma_{\rho\mu})_\beta{}^\alpha\ .
\label{AEpsYLambda}
\end{eqnarray}
Next we use the supersymmetry algebra acting on $\lambda$, which is
expressed in the Slavnov--Taylor identity by the terms proportional to
$\epsilonbar\epsilon Y_\lambda$:
\begin{eqnarray}
0 & = & \dpartial{2}{\epsilonbar\partial\epsilon}\dfunc{2}{\lambda \delta Y_\lambda}
S(\Gamma) 
\\
\Rightarrow 0 & = &
{\textstyle\frac{\partial^2(sA^\mu)}{\partial\lambda\partial\epsilonbar}} 
                    \Gamma_{\epsilon Y_\lambda A^\mu}
                  + \Gamma_{\epsilonbar\epsilon Y_\lambda Y_\lambdabar}
                    \Gamma_{\lambda\lambdabar}
                  + 
{\textstyle\frac{\partial^2(s\omega^\mu)}{\partial\epsilonbar\partial\epsilon}}
                    \Gamma_{\lambda Y_\lambda \omega^\mu}
\ .
\label{LambdaAlgebra}
\end{eqnarray}
The only unknown here is the vertex function with two external sources
corresponding to an eqs.-of-motion term in the algebra
(\ref{SusyAlgebra}). Solving (\ref{LambdaAlgebra}) yields
\begin{eqnarray}
\Gamma_{\epsilonbar_\betadot\epsilon^\beta Y_\lambda{}_\gamma 
Y_\lambdabar{}^\gammadot}(p,-p)
\Gamma_{\lambda_\alpha\lambdabar_\gammadot}(-p,p)
=
\delta_\beta{}^\gamma \psl^{\betadot\alpha}\ .
\label{SymYLambda}
\end{eqnarray}
For the supersymmetry transformation of the photino into
$|\phi_{L,R}|^2$ we derive the equation
\begin{eqnarray}
0 & = & \dpartial{}{\epsilon}\dfunc{3}{\phi_L^\dagger \delta\phi_L
  \delta\lambdabar} S(\Gamma)
\\
\Rightarrow
0 & = &
{\textstyle\frac{\partial^2(sA^\mu)}{\partial\lambdabar\partial\epsilon}} 
        \Gamma_{\phi_L^\dagger\phi_L A^\mu}
+       \Gamma_{\phi_L^\dagger\phi_L \epsilon Y_\lambda}
        \Gamma_{\lambdabar\lambda}
+       \Gamma_{\phi_L^\dagger  \epsilon Y_\Psibar}
        \Gamma_{\phi_L \lambdabar \Psibar}
\ .
\label{PhiPsiLambda}
\end{eqnarray}
For $p_\lambdabar=0$ this equation may be used to determine
$\Gamma_{\phi_L \lambdabar \Psibar}$ (see
eq.~(\ref{PhiPsiLambdaCond})), for $p_\lambdabar\ne0$ it may be 
used as a symmetry condition for $\Gamma_{\phi_L^\dagger\phi_L
  \epsilon Y_\lambda}$.

Now we proceed with symmetry conditions for the supersymmetry
transformations of the matter fields. While the mass identity
(\ref{MassIdentity}) fixes the ratio of the supersymmetry
transformations $\phi\leftrightarrow\Psi$, the supersymmetry algebra
\begin{eqnarray}
0 & = & \dpartial{2}{\epsilonbar\partial\epsilon}\dfunc{2}{\phi_L
               \delta Y_{\phi_L}} S(\Gamma)
\\
\Rightarrow
0 & = &  \Gamma_{Y_{\phi_L} \epsilonbar\epsilon Y_{\phi_L^\dagger}}
         \Gamma_{\phi_L\phi_L^\dagger}
+        \Gamma_{\phi_L\epsilonbar Y_{\Psi}}
         \Gamma_{\epsilon Y_{\phi_L} \Psi}  
+
{\textstyle\frac{\partial^2(s\omega^\mu)}{\partial\epsilonbar\partial\epsilon}}
         \Gamma_{\phi_L Y_{\phi_L} \omega^\mu}
\label{Algebra}
\end{eqnarray}
fixes the product. For on-shell momentum the eqs.-of-motion term 
vanishes and (\ref{Algebra}) reduces to 
\begin{eqnarray}
2\psl_{\beta\betadot} & = & 
 \Gamma_{\phi_L\epsilonbar^\betadot Y_{\Psi}}(p,-p)
   \Gamma_{\epsilon^\beta Y_{\phi_L}
     \Psi} (-p,p) \mbox{ for } p^2=m^2 \ .
\end{eqnarray}
Solving for the individual vertex functions is best done using the 
covariant decompositions
\begin{eqnarray}
\Gamma_{\phi_L \epsilonbar^\betadot Y_{\psibar_R}{}_\alphadot}(p,-p) 
& = & -\sqrt{2}\Theta_1(p^2) m\delta^\alphadot{}_\betadot
\ ,\\
\Gamma_{\phi_L \epsilonbar^\betadot Y_{\psi_L}{}^\alpha}(p,-p)
& = & -\sqrt{2}\Theta_2(p^2) \psl_{\alpha\betadot}
\end{eqnarray}
with $\Theta_1(m^2)=\Theta_2(m^2)$ due to (\ref{MassIdentity}), 
(\ref{SymPsiPsi}). The results are the following symmetry conditions:
\begin{eqnarray}
\mbox{for } p^2 = m^2\ :&&\nonumber\\
\label{PrefactorsStart}
\Gamma_{\phi_L \epsilonbar^\betadot Y_{\psi_L}{}^\alpha}(p,-p)
& = & -\sqrt{2} \psl_{\alpha\betadot}\ \Theta
\ ,\\
\Gamma_{\phi_L \epsilonbar^\betadot Y_{\psibar_R}{}_\alphadot}(p,-p) 
& = & -\sqrt{2} m\delta^\alphadot{}_\betadot\ \Theta
\label{PsiPrefactorsEnd}
\ ,\\
\psl_{\alpha\betadot}
 \Gamma_{\psi_L{}_\alpha \epsilon^\beta Y_{\phi_L}} (p,-p) 
& = &
- m \Gamma_{\psibar_R{}^\betadot \epsilon^\beta Y_{\phi_L}} (p,-p) 
-\sqrt{2}\psl_{\beta\betadot}\ \frac{1}{\Theta}
\label{PrefactorsEnd}
\label{PhiPrefactors}
\ ,\\
\Theta & = & \lim_{p^2\to m^2}\sqrt{{Z_\psi(p^2)}/{Z_\phi(p^2)}}\ .
\end{eqnarray}
Using these results together with the gauge covariance of the
supersymmetry transformation of $\psi_L$
\begin{eqnarray}
0 & = & \dpartial{}{\epsilonbar^\betadot}\dfunc{3}{c\delta\phi_L\delta
  Y_{\psi_L}^\alpha}S(\Gamma)
\end{eqnarray} 
then yields
\begin{eqnarray}
q^\mu\Gamma_{A^\mu\phi_L\epsilonbar^\betadot
  Y_{\psi_L}^\alpha} (q,p,p^\prime) = 
\sqrt{2}eQ_L \qsl_{\alpha\betadot} \ \Theta
\mbox{ for } p^2 = p^\prime{}^2 = m^2
\ .
\label{PhiAPrefactors}
\end{eqnarray}
Finally we determine the coefficient of the eqs.-of-motion term in the
supersymmetry algebra acting on $\psi_L$, given by
$\Gamma_{\epsilon\epsilonbar Y_{\psi_L}  Y_{\psibar_L}}$:  
\begin{eqnarray}
0 & = & \dpartial{2}{\epsilon \partial \epsilonbar}
        \dfunc{2}{\psi_L \delta Y_{\psi_L}}
  S(\Gamma)
\\
\Rightarrow 
0 & = & \Gamma_{\epsilon^\beta\epsilonbar^\betadot Y_{\psi_L}^\gamma
           Y_{\psibar_L}^\deltadot} 
        \Gamma_{\psi_L{}_\alpha \psibar_L{}_\deltadot}
  +     \Gamma_{\epsilon^\beta\epsilonbar^\betadot Y_{\psi_L}^\gamma
           Y_{\psi_R}{}_\delta}
        \Gamma_{\psi_L{}_\alpha \psi_R^\delta}
\nonumber\\&&{}
   \Gamma_{\psi_L{}_\alpha\epsilon^\beta Y_{\phi_L}}
   \Gamma_{Y_{\psi_L}^\gamma \epsilonbar^\betadot \phi_L}
+  \Gamma_{\psi_L{}_\alpha\epsilonbar^\betadot Y_{\phi_R^\dagger}}
   \Gamma_{Y_{\psi_L}^\gamma \epsilon^\beta \phi_R^\dagger}
\nonumber\\&&{}
 - 2\psl_{\beta\betadot}\delta_{\alpha}^\gamma \ .
\end{eqnarray}
Since all other vertex functions of dimension $\le4$ have already been
fixed, this identity can be viewed as a symmetry condition for  
$\Gamma_{\epsilon\epsilonbar Y_{\psi_L} Y_{\psibar_L}}$.

\subsection{Physical conditions}
\label{SecPhysCond}

In addition to the mass equality from subsection \ref{SecMass} here we
derive further symmetry conditions for physical vertex
functions. Thereby we make use of the conditions derived in
sec.~\ref{SecHigherOrderSusy}, expressing the higher order
modifications to the supersymmetry transformations, and of the
requirements (\ref{GEQ}), (\ref{WI}) that there are no higher order
corrections to gauge transformations.

Due to supersymmetry, the photon and photino self energies are related:
\begin{eqnarray}
0 & = & \dpartial{}{\epsilon}\dfunc{2}{A^\rho \delta\lambdabar} S(\Gamma)
\\
\Rightarrow
0 & = &
{\textstyle\frac{\partial^2(sA^\mu)}{\partial\lambdabar\partial\epsilon}} 
        \Gamma_{A^\rho A^\mu}
+       \Gamma_{A^\rho \epsilon Y_\lambda}
        \Gamma_{\lambdabar \lambda}
\ .
\end{eqnarray}
The prefactor $\Gamma_{A^\rho \epsilon Y_\lambda}$, expressing the
supersymmetry transformation of the 
photino, is determined to all orders by (\ref{AEpsYLambda}) and thus
\begin{eqnarray}
\sigma^\mu_{\beta\alphadot}\Gamma_{A^\rho A^\mu}(p,-p)
= -i{p}^\nu(\sigma_{\nu\rho})_\beta{}^\alpha
\Gamma_{\lambdabar^\alphadot\lambda^\alpha}(-p,p) \ .
\label{PhotonPhotino}
\end{eqnarray}
We can use this identity together with the normalization condition
(\ref{NormPhoton}) and the symmetry condition (\ref{SymYLambda}) to
get simpler conditions:
\begin{eqnarray}
\Gamma_{\lambdabar^\alphadot\lambda^\alpha}(-p,p) 
& = & \psl_{\alpha\alphadot} \mbox{ for } p^2=0 \ ,
\\
\Gamma_{\epsilonbar_\betadot\epsilon^\beta Y_\lambda{}_\gamma 
Y_\lambdabar{}^\gammadot}(p,-p) & = & \delta^\betadot{}_\gammadot
\delta_\beta{}^\gamma  \mbox{ for } p^2=0 \ .
\end{eqnarray}
Using suitable derivatives of the Ward identity (\ref{WI}) we find
that gauge invariance restricts the remaining power-counting
renormalizable photon and photino interactions: 
\begin{eqnarray}
0 & = & p^\mu \Gamma_{A^\rho A^\mu}(-p,p)
\ ,\\
0 & = & p^\mu \Gamma_{A^\rho A^\sigma A^\mu}(p^\prime,-p-p^\prime,p)
\ ,\\
0 & = & p^\mu \Gamma_{A^\rho A^\sigma A^\nu A^\mu}
 (p^\prime,p'',-p-p'-p'',p)
\ ,\\
0 & = & p^\mu \Gamma_{\lambda \lambdabar A^\mu}(p^\prime,-p-p^\prime,p)
\ .
\end{eqnarray}

Similarly, gauge invariance (\ref{WI}) yields symmetry conditions for
the photon--matter interactions, in particular 
\begin{eqnarray}
q^\mu \Gamma_{\Psi\Psibar A^\mu} (p, p^\prime,q)
& = & -eQ_L \left(\Gamma_{\Psi\Psibar}(-p^\prime,p^\prime)
            -  \Gamma_{\Psi\Psibar}(p,-p) \right)
\ .
\end{eqnarray}
Taking the derivative with respect to $q^\mu$ at $q=0$ and the limit
$p^2\to m^2$ and multiplying with spinors satisfying the Dirac
equation $(\psl-m) u(p) =0$, yields the Thomson-limit condition
\begin{eqnarray}
\bar{u}(p) Z_\Psi \Gamma_{\Psi \Psibar A^\mu}(p,-p,0) u(p) & = &
\bar{u}(p) \left(-eQ_L\gamma_\mu\right)u(p)
\nonumber\\&&{}\qquad\qquad\mbox{ for } {p^2 = m^2}
\ .
\label{PsiPsiGamma}
\end{eqnarray}
Thomson-limit conditions for the photon--selectron interactions may be
obtained in the same 
way:
\begin{eqnarray}
Z_\phi \Gamma_{\phi_L \phi_L^\dagger A^\mu}(p,-p,0) & = & 
-2eQ_L p_\mu
\quad\mbox{ for } {p^2 = m^2}
\label{PhiPhiA}
\ ,\\
Z_\phi \Gamma_{\phi_L \phi_L^\dagger A^\nu A^\mu}(p,-p,0,0) & = & 
2(eQ_L)^2 g_{\mu\nu}
\quad\mbox{ for } {p^2 = m^2}
\ .
\end{eqnarray}
The functions $Z_\Psi(p^2), Z_\phi(p^2)$ have been defined in
eqs.~(\ref{ZFactorsStart}), (\ref{ZFactorsEnd}). For brevity the
momentum arguments have been suppressed.
Instead of gauge invariance, supersymmetry is responsible for a
Thomson-limit condition for the photino--matter interaction. Using
(\ref{PhiPsiLambda}) for $p_\lambdabar=0$ together with  
(\ref{PhiPhiA}) and (\ref{PrefactorsStart}),
(\ref{PsiPrefactorsEnd})  it can be derived either in terms of
2-spinors: 
\begin{eqnarray}
\sqrt{Z_\phi Z_\Psi}
\Bigl(
\Gamma_{\phi_L^\dagger\psi_L{}_\alpha\lambda^\beta}(-p,p,0) \psl_{\alpha\betadot}
&&\nonumber\\{}
+\Gamma_{\phi_L^\dagger\psibar_R^\alphadot\lambda^\beta}(-p,p,0)
   m\delta^\alphadot_\betadot
\Bigr)
& = & 
-i\sqrt{2} eQ_L\psl_{\beta\betadot}
\quad\mbox{ for } p^2 = m^2\ ,
\label{PhiPsiLambdaCond}
\end{eqnarray}
or of 4-spinors:
\begin{eqnarray}
\sqrt{Z_\phi Z_\Psi} 
\Gamma_{\phi_L^\dagger \Psi \overline{\tilde{\gamma}}}(-p,p,0)
u(p)
& = & 
-\sqrt{2}eQ_L P_L u(p)
\quad\mbox{ for } p^2 = m^2\ .
\end{eqnarray}
The remaining power-counting renormalizable interactions are the
four-scalar interactions. Supersymmetry relates them to the
photino--matter interaction and thus to the gauge coupling in the
following way:
\begin{eqnarray}
0 & = & \dpartial{}{\epsilon}\dfunc{4}{\phi_L^\dagger \delta\phi_L \delta\phi_L^\dagger
  \delta\psi_L} S(\Gamma)
\\
\Rightarrow
0 & = & 2\Gamma_{\phi_L^\dagger \phi_L \epsilon Y_\lambda}
        \Gamma_{\phi_L^\dagger \psi_L \lambda}
+       2\Gamma_{\phi_L \phi_L^\dagger \psi_L \epsilon Y_{\phi_L}}  
        \Gamma_{\phi_L^\dagger \phi_L}
\nonumber\\&&{}
+       \Gamma_{\phi_L^\dagger \phi_L^\dagger \psi_L \epsilon Y_{\phi_L^\dagger}}  
        \Gamma_{\phi_L \phi_L^\dagger}
+       \Gamma_{\psi_L \epsilon Y_{\phi_L}}
        \Gamma_{\phi_L^\dagger \phi_L \phi_L^\dagger \phi_L}
\nonumber\\&&{}
+       \Gamma_{\phi_L^\dagger \phi_L \phi_L^\dagger  \epsilon  Y_{\Psibar}}
        \Gamma_{\psi_L \Psibar}
+       2\Gamma_{\phi_L^\dagger  \epsilon Y_\Psibar}
        \Gamma_{\phi_L^\dagger \phi_L \psi_L \Psibar}
\ ,
\end{eqnarray}
\begin{eqnarray}
0 & = & \dpartial{}{\epsilon}\dfunc{4}{\phi_R^\dagger \delta\phi_R \delta\phi_L^\dagger
  \delta\psi_L} S(\Gamma)
\\
\Rightarrow
0 & = & \Gamma_{\phi_R^\dagger \phi_R \epsilon Y_\lambda}
        \Gamma_{\phi_L^\dagger \psi_L \lambda}
+       \Gamma_{\phi_R \phi_R^\dagger \psi_L \epsilon Y_{\phi_L}}  
        \Gamma_{\phi_L^\dagger \phi_L}
+       \Gamma_{\phi_R \phi_L^\dagger \psi_L \epsilon Y_{\phi_R}}  
        \Gamma_{\phi_R^\dagger \phi_R}
\nonumber\\&&{}
+       \Gamma_{\phi_L^\dagger \phi_R^\dagger \psi_L \epsilon Y_{\phi_R^\dagger}}  
        \Gamma_{\phi_R \phi_R^\dagger}
+       \Gamma_{\psi_L \epsilon Y_{\phi_L}}
        \Gamma_{\phi_R^\dagger \phi_R \phi_L^\dagger \phi_L}
\nonumber\\&&{}
+       \Gamma_{\phi_R^\dagger \phi_R \phi_L^\dagger  \epsilon  Y_{\Psibar}}
        \Gamma_{\psi_L \Psibar}
+       \Gamma_{\phi_L^\dagger  \epsilon Y_\Psibar}
        \Gamma_{\phi_R^\dagger \phi_R \psi_L \Psibar}
+       \Gamma_{\phi_R^\dagger  \epsilon Y_\Psi}
        \Gamma_{\phi_L^\dagger \phi_R \psi_L \Psi}
\ ,
\end{eqnarray}
\begin{eqnarray}
0 & = & \dpartial{}{\epsilon}\dfunc{4}{\phi_L \delta\phi_R \delta\phi_R
  \delta\psi_L} S(\Gamma)
\\
\Rightarrow
0 & = & 2\Gamma_{\phi_L \phi_R \epsilon Y_\lambdabar}
        \Gamma_{\phi_R \psi_L \lambdabar}
+       2\Gamma_{\phi_R \phi_L \psi_L \epsilon Y_{\phi_R^\dagger}}  
        \Gamma_{\phi_R \phi_R^\dagger}
+       \Gamma_{\phi_R \phi_R \psi_L \epsilon Y_{\phi_L^\dagger}}  
        \Gamma_{\phi_L \phi_L^\dagger}
\nonumber\\&&{}
+       \Gamma_{\psi_L \epsilon Y_{\phi_L}}
        \Gamma_{\phi_L \phi_R \phi_R \phi_L}
+       \Gamma_{\phi_L \phi_R \phi_R  \epsilon  Y_{\Psibar}}
        \Gamma_{\psi_L \Psibar}
\ .
\end{eqnarray}
The momentum arguments in these terms are dropped (see explanation at
the beginning of this section). The factors 2 in front of several
terms imply symmetrization with respect to the momenta of the two
$\phi_L^\dagger$ and $\phi_R$ fields, respectively.
These equations constitute symmetry conditions for
$\Gamma_{\phi_L^\dagger \phi_L \phi_L^\dagger \phi_L}$, 
$\Gamma_{\phi_R^\dagger \phi_R \phi_L^\dagger \phi_L}$ and
$\Gamma_{\phi_R \phi_R \phi_L \phi_L}$,
since these are the only power-counting renormalizable vertex
functions not yet determined.

\subsection{Collection of all symmetry and normalization conditions}
\label{SecSymSummary}
We now list all symmetry and normalization conditions for an easy
reference and to make transparent the similarity in their mathematical
structure. Taking into account also eqs.~(\ref{Nilpotency}), (\ref{GEQ})
and the manifest symmetries there is a condition for each vertex
function corresponding to a power-counting renormalizable interaction.

\paragraph{Photon and photino only:}
\begin{eqnarray}
\mathbox{7cm}{\lim_{p^2\to0}\frac{1}{p^2}\Gamma_{A^\mu
  A^\nu}(-p,p)|_{g_{\mu\nu}-\rm part}} & = & -g_{\mu\nu}
\label{PhotonRes}
\ ,\\
\mathbox{7cm}{\Gamma_{\lambdabar^\alphadot\lambda^\alpha}(-p,p) }
& = & \psl_{\alpha\alphadot} \mbox{ for } p^2=0 
\label{PhotinoRes}
\ ,\\
\mathbox{7cm}{p^\mu \Gamma_{A^\rho A^\mu}(-p,p)} & = & 0 
\label{PhotonTransversality}
\ ,\\
\mathbox{7cm}{p^\mu \Gamma_{A^\rho A^\sigma A^\mu}
   (p^\prime,-p-p^\prime,p)} & = & 0
\ ,\\
\mathbox{7cm}{p^\mu \Gamma_{A^\rho A^\sigma A^\nu A^\mu}
 (p^\prime,p'',-p-p'-p'',p)} & = & 0
\ ,\\
\mathbox{7cm}{p^\mu \Gamma_{\lambda \lambdabar A^\mu}
   (p^\prime,-p-p^\prime,p)} & = & 0
\ ,
\end{eqnarray}
\paragraph{Interactions involving matter fields:}
\begin{eqnarray}
\label{MassStart}
\label{MatterStart}
\mathbox{6cm}{\Gamma_{\phi_L\phi_L^\dagger}(-p,p)} & = & 0 
\quad\mbox{ for } p^2=m^2
\ ,\\
\mathbox{6cm}{\Gamma_{\Psi\Psibar}(p,-p) u(p)} & = & 0 
\quad\mbox{ for } p^2=m^2 
\label{MassEnd}
\ ,\\
\mathbox{6cm}{\frac{\partial}{\partial p^2} \Gamma_{\phi_L\phi_L^\dagger}(-p,p)} 
 & = & 1 
\quad\mbox{ for } p^2=\kappa^2
\ ,\\
\mathbox{6cm}{\Gamma_V(p^2) + 2m^2(\Gamma_V^\prime(p^2) +\Gamma_S^\prime(p^2))} & = &
1  
\quad\mbox{ for } p^2=\kappa^2
\label{MatterEnd}
\ ,
\end{eqnarray}
\begin{eqnarray}
\label{ChargeStart}
\mathbox{6cm}{Z_\phi \Gamma_{\phi_L \phi_L^\dagger A^\mu}(p,-p,0)} & = & 
\mathbox{2cm}{-2eQ_L p_\mu}
\quad\mbox{ for } {p^2 = m^2}
\ ,\\
\mathbox{6cm}{Z_\phi \Gamma_{\phi_L \phi_L^\dagger A^\nu A^\mu}(p,-p,0,0)}& = & 
\mathbox{2cm}{2(eQ_L)^2 g_{\mu\nu}}
\quad\mbox{ for } {p^2 = m^2}
\ ,\\
\mathbox{6cm}{\bar{u}(p) Z_\Psi \Gamma_{\Psi \Psibar A^\mu}(p,-p,0) u(p)} & = &
\bar{u}(p) \left(-eQ_L\gamma_\mu\right)u(p)
\nonumber\\&&{}\mathbox{2cm}{}\quad\mbox{ for } {p^2 = m^2}
\ ,\\
\mathbox{6cm}{\sqrt{Z_\phi Z_\Psi} 
\Gamma_{\phi_L^\dagger \Psi \overline{\tilde{\gamma}}}(-p,p,0)
u(p)}
& = & 
-\sqrt{2}eQ_L P_L u(p)
\nonumber\\&&{}\mathbox{2cm}{}\quad\mbox{ for } p^2 = m^2
\label{ChargeEnd}
\ ,
\end{eqnarray}
\begin{eqnarray}
0 & = & 2\Gamma_{\phi_L^\dagger \phi_L \epsilon Y_\lambda}
        \Gamma_{\phi_L^\dagger \psi_L \lambda}
+       2\Gamma_{\phi_L \phi_L^\dagger \psi_L \epsilon Y_{\phi_L}}  
        \Gamma_{\phi_L^\dagger \phi_L}
\nonumber\\&&{}
+       \Gamma_{\phi_L^\dagger \phi_L^\dagger \psi_L \epsilon Y_{\phi_L^\dagger}}  
        \Gamma_{\phi_L \phi_L^\dagger}
+       \Gamma_{\psi_L \epsilon Y_{\phi_L}}
        \Gamma_{\phi_L^\dagger \phi_L \phi_L^\dagger \phi_L}
\nonumber\\&&{}
+       \Gamma_{\phi_L^\dagger \phi_L \phi_L^\dagger  \epsilon  Y_{\Psibar}}
        \Gamma_{\psi_L \Psibar}
+       2\Gamma_{\phi_L^\dagger  \epsilon Y_\Psibar}
        \Gamma_{\phi_L^\dagger \phi_L \psi_L \Psibar}
\ ,\\
0 & = & \Gamma_{\phi_R^\dagger \phi_R \epsilon Y_\lambda}
        \Gamma_{\phi_L^\dagger \psi_L \lambda}
+       \Gamma_{\phi_R \phi_R^\dagger \psi_L \epsilon Y_{\phi_L}}  
        \Gamma_{\phi_L^\dagger \phi_L}
+       \Gamma_{\phi_R \phi_L^\dagger \psi_L \epsilon Y_{\phi_R}}  
        \Gamma_{\phi_R^\dagger \phi_R}
\nonumber\\&&{}
+       \Gamma_{\phi_L^\dagger \phi_R^\dagger \psi_L \epsilon Y_{\phi_R^\dagger}}  
        \Gamma_{\phi_R \phi_R^\dagger}
+       \Gamma_{\psi_L \epsilon Y_{\phi_L}}
        \Gamma_{\phi_R^\dagger \phi_R \phi_L^\dagger \phi_L}
\nonumber\\&&{}
+       \Gamma_{\phi_R^\dagger \phi_R \phi_L^\dagger  \epsilon  Y_{\Psibar}}
        \Gamma_{\psi_L \Psibar}
+       \Gamma_{\phi_L^\dagger  \epsilon Y_\Psibar}
        \Gamma_{\phi_R^\dagger \phi_R \psi_L \Psibar}
+       \Gamma_{\phi_R^\dagger  \epsilon Y_\Psi}
        \Gamma_{\phi_L^\dagger \phi_R \psi_L \Psi}
\ ,
\\
0 & = & 2\Gamma_{\phi_L \phi_R \epsilon Y_\lambdabar}
        \Gamma_{\phi_R \psi_L \lambdabar}
+       2\Gamma_{\phi_R \phi_L \psi_L \epsilon Y_{\phi_R^\dagger}}  
        \Gamma_{\phi_R \phi_R^\dagger}
+       \Gamma_{\phi_R \phi_R \psi_L \epsilon Y_{\phi_L^\dagger}}  
        \Gamma_{\phi_L \phi_L^\dagger}
\nonumber\\&&{}
+       \Gamma_{\psi_L \epsilon Y_{\phi_L}}
        \Gamma_{\phi_L \phi_R \phi_R \phi_L}
+       \Gamma_{\phi_L \phi_R \phi_R  \epsilon  Y_{\Psibar}}
        \Gamma_{\psi_L \Psibar}
\ ,
\end{eqnarray}

\paragraph{Interactions involving ghost fields:}
\begin{eqnarray}
\mathbox{4cm}{\Gamma_{A^\mu \epsilon^\beta Y_\lambda{}_\alpha}(p,-p)}
& = & 
p^\rho (\sigma_{\rho\mu})_\beta{}^\alpha
\ ,\\
\mathbox{4cm}{\Gamma_{\epsilonbar_\betadot\epsilon^\beta Y_\lambda{}_\gamma 
Y_\lambdabar{}^\gammadot}(p,-p)}
& = & 
\mathbox{3cm}{\delta^\betadot{}_\gammadot
\delta_\beta{}^\gamma}  \mbox{ for } p^2=0 
\ ,\\
\mathbox{4cm}{-   \Gamma_{\phi_L^\dagger\phi_L \epsilon Y_\lambda}
        \Gamma_{\lambdabar\lambda}}
 & = & \Gamma_{\lambdabar\epsilon Y_{A^\mu}} 
        \Gamma_{\phi_L^\dagger\phi_L A^\mu}
+       \Gamma_{\phi_L^\dagger  \epsilon Y_\Psibar}
        \Gamma_{\phi_L \lambdabar \Psibar}
\ ,
\end{eqnarray}
\begin{eqnarray}
\label{SusyStart}
\mathbox{4cm}{\Gamma_{\phi_L \epsilonbar^\betadot 
 Y_{\psi_L}{}^\alpha}(p,-p)}
& = & \mathbox{3cm}{-\sqrt{2} \psl_{\alpha\betadot}\ \Theta}
\mbox{ for } p^2 = m^2
\ ,\\
\mathbox{4cm}{\Gamma_{\phi_L \epsilonbar^\betadot 
 Y_{\psibar_R}{}_\alphadot}(p,-p)}
& = & \mathbox{3cm}{-\sqrt{2} m\delta^\alphadot{}_\betadot\ \Theta}
\mbox{ for } p^2 = m^2
\ ,\\
\mathbox{4cm}{\psl_{\alpha\betadot}
 \Gamma_{\psi_L{}_\alpha \epsilon^\beta Y_{\phi_L}} (p,-p)}
& = &
- m \Gamma_{\psibar_R{}^\betadot \epsilon^\beta Y_{\phi_L}} (p,-p) 
-\sqrt{2}\psl_{\beta\betadot}\ \frac{1}{\Theta}
\nonumber\\&&{}\mathbox{3cm}{}
\mbox{ for } p^2 = m^2
\ ,\\
\mathbox{4cm}{q^\mu\Gamma_{A^\mu\phi_L\epsilonbar^\betadot
  Y_{\psi_L}^\alpha} (q,p,p^\prime)} & = & 
\mathbox{3cm}{\sqrt{2}eQ_L \qsl_{\alpha\betadot} \ \Theta}
\mbox{ for } p^2 = p^\prime{}^2 = m^2
\ ,\\
0\quad
 & = & \Gamma_{\epsilon^\beta\epsilonbar^\betadot Y_{\psi_L}^\gamma
           Y_{\psibar_L}^\deltadot} 
        \Gamma_{\psi_L{}_\alpha \psibar_L{}_\deltadot}
  +     \Gamma_{\epsilon^\beta\epsilonbar^\betadot Y_{\psi_L}^\gamma
           Y_{\psi_R}{}_\delta}
        \Gamma_{\psi_L{}_\alpha \psi_R^\delta}
\nonumber\\&&{}
   \Gamma_{\psi_L{}_\alpha\epsilon^\beta Y_{\phi_L}}
   \Gamma_{Y_{\psi_L}^\gamma \epsilonbar^\betadot \phi_L}
+  \Gamma_{\psi_L{}_\alpha\epsilonbar^\betadot Y_{\phi_R^\dagger}}
   \Gamma_{Y_{\psi_L}^\gamma \epsilon^\beta \phi_R^\dagger}
\nonumber\\&&{}
 - 2\psl_{\beta\betadot}\delta_{\alpha}^\gamma 
\ .
\label{SusyEnd}
\end{eqnarray}


\section{Applications}
\label{SecApplications}
The general prescription for higher order calculations not relying on
an invariant regularization is: 
\begin{itemize}
\item Calculate the necessary loop diagrams using some arbitrary
  (preferably consistent) regularization. 
\item To every power-counting renormalizable interaction there is an
  independent counterterm.
\item For each counterterm the proper coefficient can be read off from
  one of the conditions collected in section \ref{SecSymSummary}.
\item From the considerations in section \ref{SecDefinition} we know
  that this leads uniquely to a renormalized theory respecting all
  defining symmetries.
\end{itemize}
In this section we show some sample calculations of renormalized
higher order corrections using dimensional regularization as defined
in \cite{BM77}. In particular we use $\{\gamma^\mu,\gamma^5\} =
2\hat{g}^{\mu\nu}\gamma_\nu\gamma^5$ with $\hat{g}^\mu{}_\mu = D-4$
and set $\hat{g}^{\mu\nu}=0$ only in the final results. This
regularization scheme is known to break supersymmetry. In establishing
the symmetries of the renormalized theory, the symmetry conditions we
have derived will prove to be an efficient tool, due to the 
common structure of most of them: 
\begin{eqnarray}
\Gamma_{ABC}|_{\rm on\ shell} = \Gamma_{ABC}^{\rm regularized}
+ \Gamma_{ABC}^{\rm ct} = \mbox{definite value}.
\end{eqnarray}
Non-supersymmetric counterterms in dimensional regularization have
already been calculated in the literature \cite{MSBar}. The equality
of the effective couplings to gauge bosons and gauginos we have proven
in sec. \ref{SecSymCond} as a consequence of the defining symmetry
requirements was anticipated there as a symmetry condition and used
for the determination of the counterterms. 

\subsection{Elimination of $B$}

Although for theoretical purposes the auxiliary $B$ field is useful,
it complicates practical calculations whenever we are not interested
in Green functions involving external $B$ fields. Therefore it is
convenient to eliminate $B$ by its equation of motion. Due to the
gauge condition in eq.~(\ref{GEQ}) we can write  
\begin{eqnarray}
\Gamma(B,A_\mu,\ldots) = \Gamma_{{\rm no\ } B}(A_\mu,\ldots)
 + \Gamma_{{\rm with\ } B}(B,A_\mu) \ ,
\end{eqnarray}
where the first term does not depend on $B$ and
\begin{eqnarray}
\Gamma_{{\rm with\ } B}(B,A_\mu) 
= \intx\Bigl( B \partial^\mu A_\mu + \frac{\xi}{2} B^2\Bigr) \ .
\end{eqnarray}
The solution of the equation of motion is
$B=-\frac{1}{\xi}(\partial A)$ to all orders, and one can show that the
effective action  
\begin{eqnarray}
\tilde{\Gamma}(A_\mu,\ldots) & = & \Gamma_{{\rm no\ } B}(A_\mu,\ldots)
 + \Gamma_{{\rm with\ } B}
       ({\textstyle B=-\frac{1}{\xi}(\partial A),A_\mu}) 
\nonumber\\
& = &  \Gamma_{{\rm no\ } B}(A_\mu,\ldots)
 -\frac{1}{2\xi}\intx(\partial^\mu A_\mu)^2\ ,
\end{eqnarray}
where $\tilde{\Gamma}$ does not depend on $B$, 
generates the same connected Green functions as
$\Gamma(B,A_\mu,\ldots)$. In the passage from $\Gamma$ to
$\tilde{\Gamma}$, the only vertex function that changes is
$\Gamma_{A^\mu A^\rho}$, which receives a longitudinal part.
In the rest of this section we always work with $\tilde{\Gamma}$, so
we drop the $\tilde{ }$ and denote by $\Gamma$ the effective action
without $B$. This yields 
\begin{eqnarray}
p^\mu \Gamma_{A^\rho A^\mu}(-p,p) = -\frac{1}{\xi}p^2 p_\rho
\label{PhotonPropagator}
\end{eqnarray}
instead of eq.~(\ref{PhotonTransversality}), while all other
conditions in section \ref{SecSymSummary} are unchanged.

\subsection{Photon and photino self energies}

\begin{figure}[tb]
\begin{center}
\begin{picture}(400,45)
\epsfxsize=3cm
\put(300,-10){\epsfbox{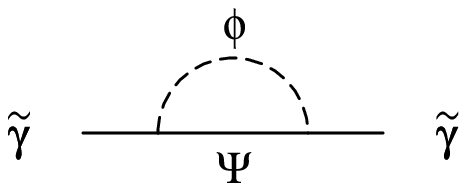}}
\epsfxsize=3cm
\put(00,0){\epsfbox{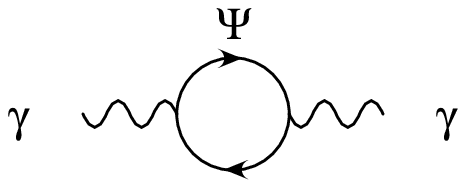}}
\epsfxsize=3cm
\put(100,0){\epsfbox{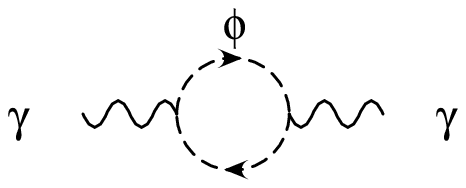}}
\epsfxsize=3cm
\put(200,0){\epsfbox{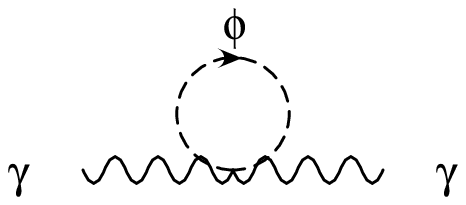}}
\end{picture}
\end{center}
\caption{One-loop diagrams contributing to the photon and
  photino self energies.} 
\label{FigPhoton}
\end{figure}

The one-loop diagrams contributing to the photon and photino self
energies are depicted in fig.~\ref{FigPhoton}. In terms of the
one-loop integrals defined in app.~\ref{App1LInt}, the results are
($\alpha=\frac{e^2}{4\pi}$)
\begin{eqnarray}
\Gamma^{\rm regularized}_{A^\mu A^\rho}(-p,p) & = & \left(-g_{\mu\rho}p^2+p_\mu
  p_\rho\right) (1+\Pi^\gamma(p^2)) - \frac{1}{\xi}p_\mu p_\rho 
\ ,\\
\Gamma^{\rm regularized}_{\lambdabar^\alphadot\lambda^\alpha}(-p,p) 
& = & \psl_{\alpha\alphadot} (1+\Pi^{\tilde{\gamma}}(p^2))
\ ,
\end{eqnarray}
where the one-loop corrections 
\begin{eqnarray}
\Pi^\gamma(p^2) = \Pi^{\tilde{\gamma}}(p^2)
= \frac{\alpha}{4\pi} 2 B_0(m^2,m^2,p^2)
\end{eqnarray}
turn out to be equal, so the identity (\ref{PhotonPhotino}) is already
satisfied at the regularized level (up to the new longitudinal part of
$\Gamma_{A^\mu A^\rho}$). To renormalize we have to define
counterterms such that the conditions (\ref{PhotonRes}),
(\ref{PhotinoRes}) are satisfied. The correct choice is
\begin{eqnarray}
\L_{\rm ct} = \delta Z_\gamma (-\frac{1}{4}F_{\mu\nu}F^{\mu\nu} +
\frac{1}{2}\overline{\tilde{\gamma}}i\gamma^\mu\partial_\mu\tilde{\gamma} 
)
\end{eqnarray}
with
\begin{eqnarray}
\delta Z_\gamma = -\Pi^\gamma(0)
\ ,
\end{eqnarray}
yielding to $\cal O(\alpha)$
\begin{eqnarray}
\Gamma_{A^\mu A^\rho}(-p,p) & = & \left(-g_{\mu\rho}p^2+p_\mu
  p_\rho\right) (1+\Pi^\gamma(p^2)+\delta Z_\gamma) - \frac{1}{\xi}p_\mu p_\rho 
\ ,\\
\Gamma_{\lambdabar^\alphadot\lambda^\alpha}(-p,p) 
& = & \psl_{\alpha\alphadot} (1+\Pi^{\gamma}+\delta Z_\gamma(p^2))
\ .
\end{eqnarray}
Note that mass and gauge fixing counterterms are not ruled out a
priori but they turn out to vanish because of the concrete form of the
regularized self energies.

\subsection{Electron and selectron self energies}

\begin{figure}[tb]
\begin{center}
\begin{picture}(400,85)
\epsfxsize=3cm
\put(200,35){\epsfbox{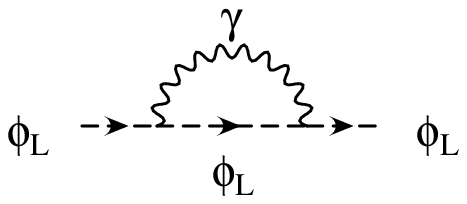}}
\epsfxsize=3cm
\put(300,35){\epsfbox{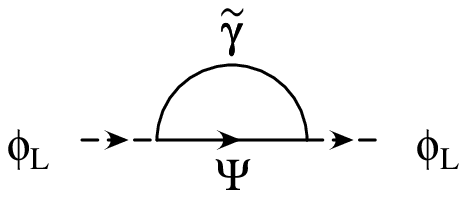}}
\epsfxsize=3cm
\put(00,35){\epsfbox{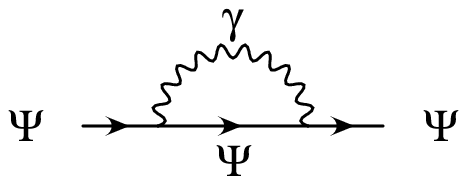}}
\epsfxsize=3cm
\put(100,35){\epsfbox{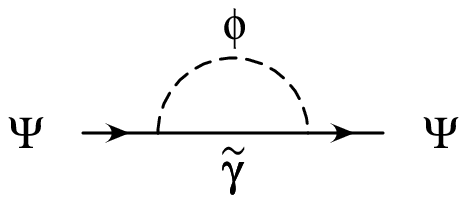}}
\epsfxsize=3cm
\put(200,-15){\epsfbox{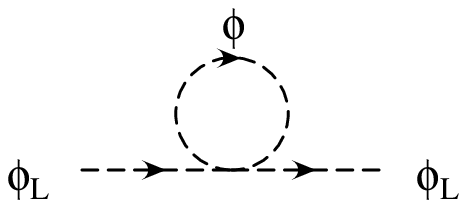}}
\epsfxsize=3cm
\put(300,-15){\epsfbox{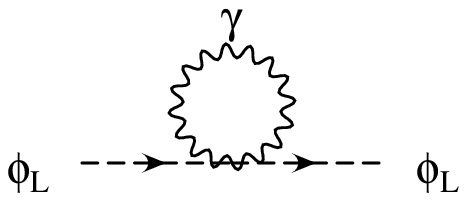}}
\end{picture}
\end{center}
\caption{One-loop diagrams contributing to the electron and
  selectron self energies.}
\label{FigMatter}
\end{figure}

The one-loop contributions to the matter self energies can be written
as follows: 
\begin{eqnarray}
\Gamma^{\rm regularized}_{\phi_L\phi_L^\dagger}(p,-p) & = & p^2-m^2 +
\Sigma_\phi(p^2) \ ,\\
\Gamma^{\rm regularized}_{\Psi\Psibar}(p,-p) & = & \psl - m +
\psl\Sigma_V(p^2) - m\Sigma_S(p^2)
\ .
\end{eqnarray}
For later purposes we also introduce the abbreviation
\begin{eqnarray}
\Sigma_\Psi'(p^2) = \Sigma_V(p^2) + 2m^2(\Sigma_V'(p^2) - \Sigma_S'(p^2))\ .
\end{eqnarray}
The contributing Feynman diagrams are shown in fig.~\ref{FigMatter}
and yield\footnote{For the rest of this section we use the gauge
  parameter $\xi=1$.}
\begin{eqnarray}
\Sigma_\phi(p^2) & = & \frac{\alpha}{4\pi}[-4m^2
B_0(0,m^2,p^2)+4(D-4)B_{22}(0,m^2,p^2)] 
\ ,\\
\Sigma_V(p^2) & = & \frac{\alpha}{4\pi}[(D-2) B_0(0,m^2,p^2) + (D-4)
B_1(0,m^2,p^2)] 
\ ,\\
\Sigma_S(p^2) & = & \frac{\alpha}{4\pi}[D B_0(0,m^2,p^2)]
\ .
\end{eqnarray}
The most general counterterms contributing to these self energies are
\begin{eqnarray}
\L_{\rm ct} &  = & \delta Z_\phi (|\partial_\mu\phi_L|^2 - m^2|\phi_L|^2 +
({}_{L\to R})) - 2m\delta m_\phi (|\phi_L|^2 + |\phi_R|^2)
\nonumber\\&&{}
+\delta Z_\Psi \Psibar (i\gamma^\mu\partial_\mu - m) \Psi  
-\delta m_\Psi \Psibar\Psi
\ .
\end{eqnarray}
For each counterterm one of the conditions
(\ref{MatterStart}--\ref{MatterEnd}) applies. Expressed in terms of
the quantities in $\L_{\rm ct}$ they read:
\begin{eqnarray}
\Sigma_\phi(m^2)-2m\delta m_\phi & = & 0
\ ,\\
m\Sigma_V(m^2)-m\Sigma_S(m^2) - \delta m_\Psi & = & 0
\ ,\\
\Sigma_\phi'(\kappa^2) + \delta Z_\phi & = & 0
\ ,\\
\Sigma_\Psi'(\kappa^2) + \delta Z_\Psi & = & 0
\ ,
\end{eqnarray}
from which the coefficients of the counterterms follow immediately:
\begin{eqnarray}
\delta m_\phi & = & \frac{\alpha}{4\pi}m \Bigl[ -2B_0(0,m^2,m^2) -\frac{2}{3}\Bigr]
\ ,\\
\delta m_\Psi & = & \frac{\alpha}{4\pi}m [ -2B_0(0,m^2,m^2) + 1]
\ ,\\
\delta Z_\phi & = & \frac{\alpha}{4\pi} \Bigl[4m^2B_0'(0,m^2,\kappa^2)-\frac{2}{3}\Bigr]
\ ,\\
\delta Z_\Psi & = & \frac{\alpha}{4\pi}
          [-2B_0(0,m^2,\kappa^2) + 4m^2B_0'(0,m^2,\kappa^2) + 1]
\ ,
\end{eqnarray}
where in the finite terms the limit $D\to4$ has been taken. 

This non-vanishing difference $\delta m_\Psi - \delta m_\phi$ is our
first encounter of a super\-sym\-metry-violating counterterm, necessary
because dimensional regularization itself breaks supersymmetry. It is
precisely this choice for the counterterms that restores
(\ref{MassStart}--\ref{MassEnd}) and thus the equality of the
renormalized masses, a necessary consequence of supersymmetry.

The different $\delta Z$ counterterms do not correspond to a
symmetry breaking, as shown in section \ref{SectionSymmetricCTs}.

\subsection{Photon and photino interactions with electron and selectron}

\begin{figure}[tb]
\begin{center}
\begin{picture}(400,135)
\epsfxsize=3cm
\put(0,-0){\epsfbox{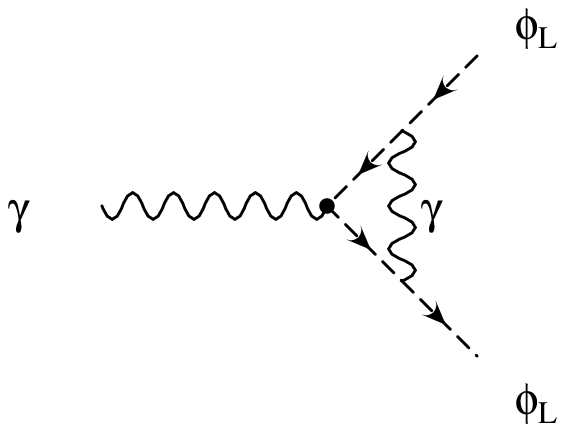}}
\epsfxsize=3cm
\put(100,-0){\epsfbox{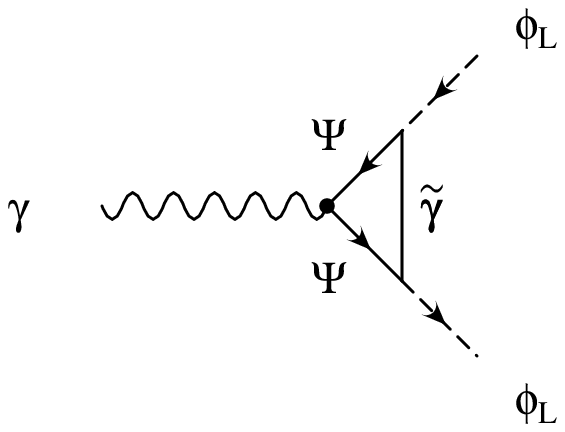}}
\epsfxsize=3cm
\put(200,0){\epsfbox{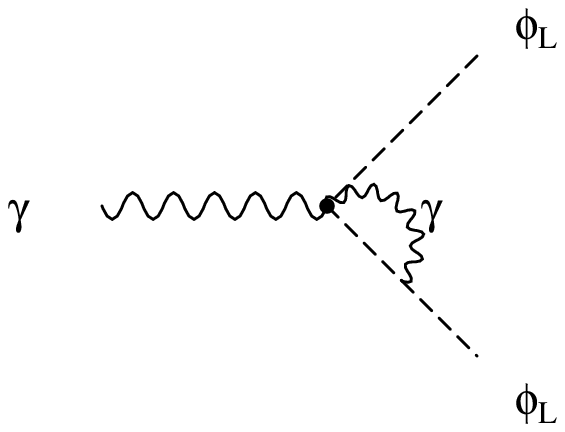}}
\epsfxsize=3cm
\put(300,0){\epsfbox{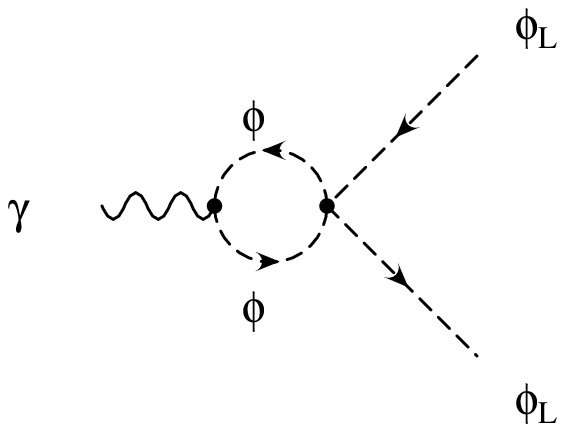}}
\epsfxsize=3cm
\put(0,65){\epsfbox{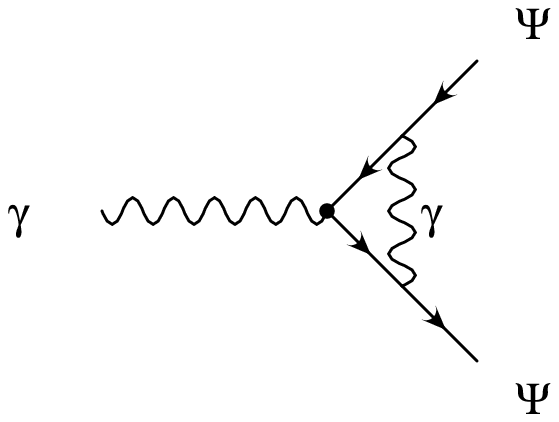}}
\epsfxsize=3cm
\put(100,65){\epsfbox{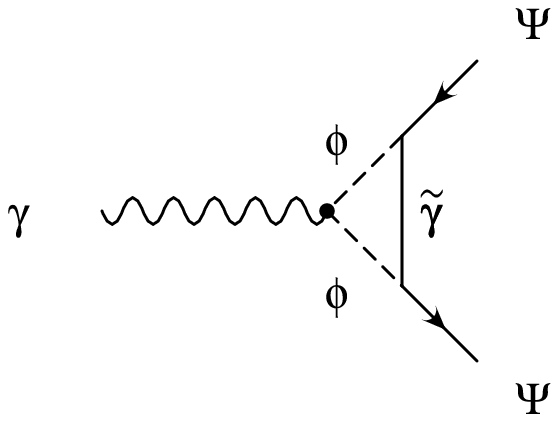}}
\epsfxsize=3cm
\put(200,65){\epsfbox{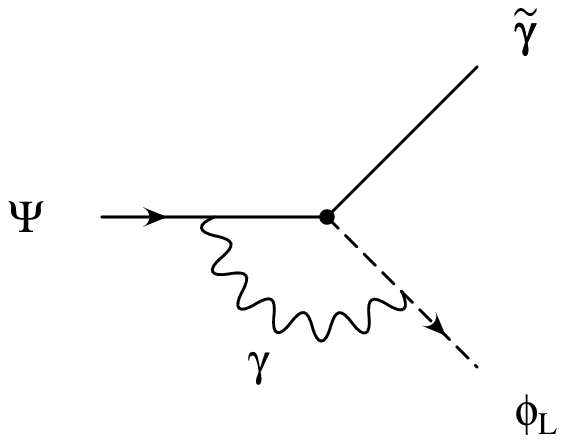}}
\epsfxsize=3cm
\put(300,65){\epsfbox{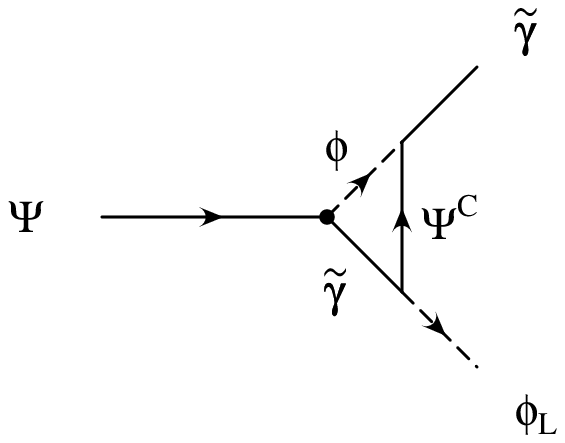}}
\end{picture}
\end{center}
\caption{One-loop vertex corrections.
 }
\label{FigCharges}
\end{figure}

We define scalar functions containing the regularized one-loop
contributions to the photon--/photino--matter interactions in the
following way:
\begin{eqnarray}
\Gamma^{\rm regularized}_{\phi_L \phi_L^\dagger A^\mu}(p,-p,0)
& = & \Lambda_{\phi\phi A}(p^2) \ (-2eQ_L p_\mu)
\ ,\\
\bar{u}(p) \Gamma^{\rm regularized}_{\Psi \Psibar A^\mu}(p,-p,0) u(p)
& = & \Lambda_{\Psi\Psi A}(p^2) \ \bar{u}(p) (-eQ_L\gamma_\mu) u(p)
\ ,\\
\Gamma^{\rm regularized}_{\phi_L^\dagger \Psi \overline{\tilde{\gamma}}}(-p,p,0) u(p)
& = & \Lambda_{\phi\Psi\tilde{\gamma}}(p^2) \ (-\sqrt{2}eQ_LP_L)u(p)
\ .
\end{eqnarray}
For each of these vertex functions there is one
independent counterterm. To make the comparison with the case of
symmetric counterterms transparent we denote them by
\begin{eqnarray}
\L_{\rm ct} & = & 
(\delta Z_\phi +\frac{1}{2}\delta Z_\gamma + \delta Z_{\phi\phi  A}) 
ieQ_LA^\mu(\phi_L^\dagger\partial_\mu\phi_L-\phi_L\partial_\mu\phi_L^\dagger)
+({}_{L\to R})
\nonumber\\&&{}
(\delta Z_\Psi +\frac{1}{2}\delta Z_\gamma + \delta Z_{\Psi\Psi A}) 
\Psibar(-eQ_L A^\mu \gamma_\mu) \Psi
\nonumber\\&&{}
(\frac{\delta Z_\phi +\delta Z_\Psi +\delta Z_\gamma}{2} + \delta
Z_{\phi\Psi\tilde{\gamma}}) (-\sqrt{2}eQ_L)(\phi_L^\dagger \overline{\tilde{\gamma}} P_L
\Psi - \phi_R \overline{\tilde{\gamma}}P_R \Psi 
 +h.c.)
\nonumber\\
\end{eqnarray}
According to section \ref{SectionSymmetricCTs} these counterterms are
symmetric if $\delta Z_{\phi\phi  A}$ = $\delta Z_{\Psi\Psi A}$
= $\delta Z_{\phi\Psi\tilde{\gamma}}$ . Their values are determined by
the conditions (\ref{ChargeStart}--\ref{ChargeEnd}). The functions
$Z_\phi,Z_\Psi$ are given in one-loop order by 
\begin{eqnarray}
Z_\phi(p^2) & = & 1 - \Sigma_\phi'(p^2) - \delta Z_\phi \ ,\\
Z_\Psi(p^2) & = & 1 - \Sigma_\Psi'(p^2) - \delta Z_\Psi \ ;
\end{eqnarray}
therefore in (\ref{ChargeStart}--\ref{ChargeEnd}) the matter field
renormalization factors $\delta Z_\phi, \delta Z_\Psi$ drop out and
the remaining conditions are 
\begin{eqnarray}
\Lambda_{\phi\phi A}(p^2) - \Sigma_\phi'(p^2) + 
\frac{1}{2}\delta Z_\gamma + \delta Z_{\phi\phi A} & = & 0
\quad\mbox{ for } {p^2 = m^2}
,\\
\Lambda_{\Psi\Psi A}(p^2) - \Sigma_\Psi'(p^2) +
\frac{1}{2}\delta Z_\gamma + \delta Z_{\Psi\Psi A} & = & 0
\quad\mbox{ for } {p^2 = m^2}
,\\
\Lambda_{\phi\Psi\tilde{\gamma}}(p^2) 
-\frac{1}{2}(\Sigma_\phi'(p^2) + \Sigma_\Psi'(p^2)) +
\frac{1}{2}\delta Z_\gamma + \delta Z_{\phi\Psi \tilde{\gamma}} & = & 0
\quad\mbox{ for } {p^2 = m^2}
\label{PhotinoCond}
.
\end{eqnarray}
Again, the counterterms can be read off easily from the
corresponding conditions once the loop diagrams shown in
fig.~\ref{FigCharges} have been calculated. Inspection of the Feynman
integrands shows that both conditions for the photon interactions
already hold at the regularized level, so we have to choose 
\begin{eqnarray}
\delta Z_{\phi\phi A} = \delta Z_{\Psi\Psi A} = 
-\frac{1}{2}\delta Z_\gamma \ .
\end{eqnarray}

Physically these conditions express the gauge invariance of the
renormalized theory, and the structure of these counterterms shows
that gauge invariance is preserved by dimensional regularization. 

The one-loop correction to the photino interaction is given by
\begin{eqnarray}
\Lambda_{\phi\Psi\tilde{\gamma}}(p^2) 
& = & \frac{\alpha}{4\pi} [B_0(0,m^2,p^2) + 4m^2(C_0+C_{11}) 
      + {\cal O}(p^2-m^2) ]
\end{eqnarray}
with $C_{ij}=C_{ij}(0,m^2,m^2,p^2,0,p^2)$, and the derivatives of the
matter self energies are
\begin{eqnarray}
\Sigma_\phi'(p^2) & = & \frac{\alpha}{4\pi} 
                        \Bigl[-4m^2B_0'(0,m^2,p^2)+\frac{2}{3}\Bigr]
\ ,\\
\Sigma_\Psi'(p^2) & = & \frac{\alpha}{4\pi}
          [2B_0(0,m^2,p^2) - 4m^2B_0'(0,m^2,p^2) - 1]
\ .
\end{eqnarray}
Using $B_0' = -C_0-C_{11}$ shows that the correct choice for the
counterterm is
\begin{eqnarray}
\delta Z_{\phi\Psi\tilde{\gamma}} = - \frac{1}{2}\delta Z_\gamma
-\frac{1}{6}\, \frac{\alpha}{4\pi}\ .
\end{eqnarray}

This result exhibits three important aspects. First, in
eq.~(\ref{PhotinoCond}) the non-local terms cancel. This is a
regularization-independent fact due to the supersymmetry. Second, on
the dimensionally regularized level there is a local violation of
eq.~(\ref{PhotinoCond}). This supersymmetry breaking has to
be cancelled choosing the charge counterterm  
$\delta Z_{\phi\Psi\tilde{\gamma}}$ different from the charge
counterterms for the photon interactions. Physically these
non-supersymmetric counterterms lead uniquely to charge universality
in the renormalized theory as required by
eqs.~(\ref{ChargeStart}--\ref{ChargeEnd}).
Third, obviously the determination of this counterterm 
$\delta Z_{\phi\Psi\tilde{\gamma}}$ is just as straightforward as the
determination of the charge counterterms for the photon interactions
before, in spite of the supersymmetry breaking. The reason is that the
main work has already been done in the derivation of the corresponding
symmetry condition.

\label{ErrorExample}
The photino--matter interaction also constitutes an example where a naive
one-loop calculation can lead to a large numerical error. Naively one
might think that the required symmetries restrict the counterterms to
those of section \ref{SectionSymmetricCTs} corresponding to field and
parameter renormalization. According to this line of reasoning one
would ignore the effects of the regularization and choose 
$\delta Z_{\phi\Psi\tilde{\gamma}} = \delta Z_{\phi\phi A} = \delta
Z_{\Psi\Psi A}$. In this section we have shown that for dimensional
regularization this amounts to forgetting the necessary term $(-\frac{1}{6}
\frac{\alpha}{4\pi})$ and spoiling charge universality and thus
supersymmetry of the renormalized theory. Since all contributions to
$\Lambda_{\phi\Psi\tilde{\gamma}}(p^2)$ are basically of the order
$\frac{\alpha}{4\pi}$, the numerical error in the renormalized one-loop
correction to the photino--electron--selectron interaction is in
general quite sizeable.

\subsection{Supersymmetry transformations at one loop}
\label{SecSusy1Loop}

\begin{figure}[tb]
\begin{center}
\begin{picture}(400,55)
\epsfxsize=3cm
\put(0,-10){\epsfbox{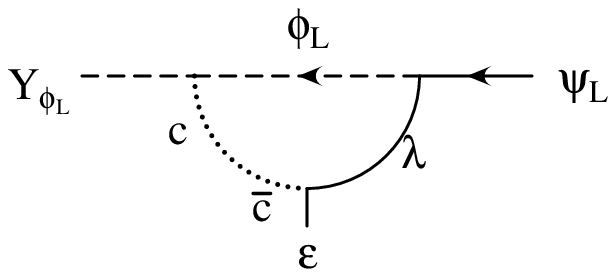}}
\epsfxsize=3cm
\put(100,-10){\epsfbox{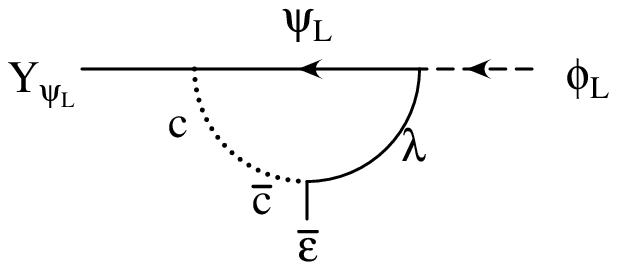}}
\epsfxsize=3cm
\put(200,-10){\epsfbox{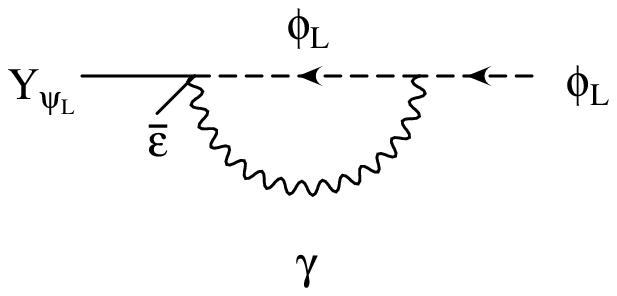}}
\epsfxsize=3cm
\put(300,-10){\epsfbox{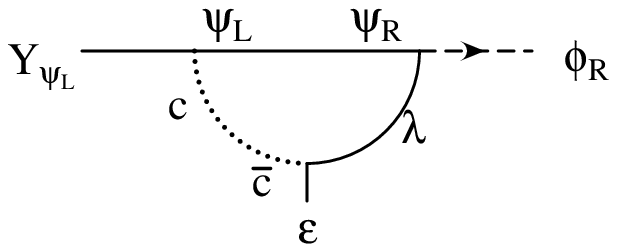}}
\end{picture}
\end{center}
\caption{One-loop contributions to the supersymmetry
  transformations of $\phi_L$ and $\psi_L$.
 }
\label{FigSusy}
\end{figure}

The Slavnov--Taylor identity may be rewritten in the form of an
invariance relation ($\varphi_i'$ runs over the linearly transforming
fields including the global ghosts, $\varphi_i, Y_i$ over the
non-linearly transforming fields and the corresponding external
fields):
\begin{eqnarray}
\Gamma(\varphi_i' + \theta s_\Gamma\varphi_i',
       \varphi_i + \theta s_\Gamma \varphi_i, Y_i) & = & 
\Gamma(\varphi_i', \varphi_i, Y_i) \ ,
\end{eqnarray}
where $\theta$ is an infinitesimal fermionic parameter and $s_\Gamma$
is the quantum analogue to the classical BRS operator:
\begin{eqnarray}
s_\Gamma\varphi_i' & = & s\varphi_i'
\ ,\\
s_\Gamma\varphi_i & = & \dg{Y_i} = \langle s_{\Gamma_{\rm cl}} \varphi_i \rangle_J
\ ,\\
s_{\Gamma_{\rm cl}}\varphi_i & = & s\varphi_i + \frac{\delta\Gamma_{\rm
    bil}}{\delta Y_i}\ .
\end{eqnarray}
$s_\Gamma\varphi_i$ is equal to the expectation value of the composite
operator $s_{\Gamma_{\rm cl}}\varphi_i$ in the presence of sources 
$J  =  -\dg{\varphi}$. 
Thus $s_\Gamma \varphi_i$ --- and equivalently the vertex functions
involving an external $Y_i$ --- contain quantum corrections to the BRS
transformations. These quantum corrections can be non-trivial but are
constrained by eqs.~(\ref{Nilpotency}), (\ref{GEQ}). 

We focus now on the transformation of the electron and selectron
fields as particular examples:
\begin{eqnarray}
s_\Gamma \phi_L(x) & = & -ieQ_L c(x)\,\phi_L(x)
  - i\omega^\nu\partial_\nu \phi_L(x)
\nonumber\\&&{}
-\int d^4y\ \epsilon^\beta\, \psi_L{}_\alpha(y) 
\ \Gamma_{\psi_L{}_\alpha\epsilon^\beta Y_{\phi_L}}(y,x)
\nonumber\\&&{}
-\int d^4y \ \epsilon^\beta\, \psibar_R{}^\alphadot(y) 
\ \Gamma_{\psibar_R{}^\alphadot\epsilon^\beta Y_{\phi_L}}(y,x) + \ldots
\ ,\\
s_\Gamma \psi_L^\alpha(x) & = & -ieQ_L c(x)\,\psi_L^\alpha(x)
-i\omega^\nu\partial_\nu \psi_L^\alpha(x)
\nonumber\\&&{}
+ \int d^4y\ 
  \epsilonbar_\betadot\, \phi_L(y) 
  \  \Gamma_{\phi_L\epsilonbar_\betadot Y_{\psi_L}{}_\alpha}(y,x)
\nonumber\\&&{}
+ \int d^4y\ 
  \epsilon^\beta\,  \phi_R^\dagger(y)
  \  \Gamma_{\phi_R^\dagger \epsilon^\beta Y_{\psi_L{}_\alpha}}(y,x)
+ \ldots\ ,
\end{eqnarray}
where the dots denote terms involving higher powers of the fields.
So the renormalized supersymmetry transformations
$\phi\leftrightarrow\psi$ are governed by vertex functions of the type 
$\Gamma_{\psi\epsilon Y_{\phi}}$ and $\Gamma_{\phi \epsilon
  Y_\psi}$.

At one-loop order these vertex functions are given by the Feynman
diagrams displayed in fig.~\ref{FigSusy} and by the counterterms 
determined through eqs.~(\ref{SusyStart}--\ref{SusyEnd}) with
$\Theta=1+\frac{\alpha}{4\pi}(B_0(0,m^2,\kappa^2)-B_0(0,m^2,m^2))$.
In momentum space the results are ($B_0 = B_0(0,m^2,p^2)$)
\begin{eqnarray}
\Gamma_{\psi_L{}_\alpha\epsilon^\beta Y_{\phi_L}} & = & 
- \sqrt{2}\delta_\beta{}^\alpha
      \left[1+\frac{\alpha}{4\pi}B_0+\frac{1}{2}\left(\delta Z_\psi
       - \delta Z_\phi -\frac{5}{3}\,\frac{\alpha}{4\pi}\right)\right]
\ ,\\
\Gamma_{\psibar_R{}^\alphadot\epsilon^\beta Y_{\phi_L}} & = & 0
\ ,\\
\Gamma_{\phi_L\epsilonbar_\betadot Y_{\psi_L}{}_\alpha} & = & 
-\sqrt{2}\psl_{\phi_L}^{\betadot\alpha}
      \left[1-\frac{\alpha}{4\pi}B_0-\frac{1}{2}\left(\delta Z_\psi
       - \delta Z_\phi -\frac{5}{3}\,\frac{\alpha}{4\pi}\right)\right]
\ ,\\
\Gamma_{\phi_R^\dagger \epsilon^\beta Y_{\psi_L{}_\alpha}} & = & 
-\sqrt{2}m\delta_{\beta}{}^\alpha
\nonumber\\
&\times&{}
      \left[1+\frac{\alpha}{4\pi}B_0-\frac{1}{2}\left(\delta Z_\psi
       - \delta Z_\phi -\frac{5}{3}\,\frac{\alpha}{4\pi}\right)
       +\frac{\delta m_\phi}{m}+\frac{2}{3}\,\frac{\alpha}{4\pi}\right] 
\, .
\end{eqnarray}
Again, non-invariant counterterms are necessary.

These results show that in one-loop order the supersymmetry
transformations are modified by non-local terms. One reason for this
modification is the non-linearity of the BRS transformations
permitting all the vertices involving $Y$ fields in
fig.~\ref{FigSusy}. Another reason can be traced back to the gauge
fixing fermion $F=\bar{c}( \partial^\mu A_\mu + \frac{\xi}{2}
B)$. Since $F$ breaks supersymmetry, there are terms in
$s_{\Gamma_{\rm cl}}F$ involving the $\epsilon$ ghosts, in particular
the $\bar{c}\epsilon\lambda$ vertices appearing in three of the graphs
in fig.~\ref{FigSusy}. These supersymmetry transformations are
related to physical vertex functions by identities such as
eq.~(\ref{MassIdentity}), (\ref{PhiPsiLambda})
expressing non-trivial relations among self energies and vertex
corrections. 

\subsection{Summary of counterterms}

We had to use non-invariant counterterms in many of the vertex
functions we calculated. However, one should note that the separation 
$\Gamma_{\rm ct}=\Gamma_{\rm sym}+\Gamma_{\rm non-inv}$ is not unique.
The simplest expression for $\Gamma_{\rm non-inv}$ is obtained using
special renormalization constants in $\Gamma_{\rm sym}$ as given by
eq.~(\ref{SymCT}). If one uses
$(\delta Z_\phi+\frac{2}{3}\frac{\alpha}{4\pi})$, $(\delta
Z_\Psi-\frac{\alpha}{4\pi})$ as field renormalization constants
instead of $\delta Z_\phi$, 
$\delta Z_\Psi$,  and the mass counterterm $m\delta Z_m = (\delta
m_\phi+\frac{2}{3}\frac{\alpha}{4\pi})$, then the non-invariant 
counterterms are confined to the matter self energies and the
photon interactions:
\begin{eqnarray}
\Gamma_{\rm n.i.} & = & \intx\ \frac{\alpha}{4\pi}\left(
           \Psibar(i\Dsl-2m)\Psi
-\frac{2}{3}|D_\mu\phi_L|^2+2m^2|\phi_L|^2+(_{L\to R})\right).
\end{eqnarray}


 \section{Conclusions}

In this article we  have constructed the Green functions of SQED in the
Wess--Zumino gauge from the Slavnov--Taylor identity without referring
to the existence of an invariant scheme. The Slavnov--Taylor identity
expresses gauge invariance, supersymmetry and translational invariance
in a single symmetry identity. For its formulation one has to
introduce several unphysical fields, namely the Faddeev--Popov ghost
$c$, global ghosts $\epsilon,\epsilonbar,\omega^\mu$ and sources $Y_i$
for all non-linear BRS transformations. The Slavnov--Taylor identity
is a complicated non-linear equation involving Green functions with
physical and unphysical fields.

We have evaluated this identity and have derived simple symmetry
conditions that resemble the normalization conditions in their
mathematical structure. These symmetry conditions constitute exact
physical statements that are valid to all orders and express lucidly
the various aspects of the symmetries. Two important examples are the
equality of the electron and selectron masses and the charge
universality in the photon and photino interactions with electron and
selectron. These are thus proven exclusively in the Wess--Zumino gauge
without using superspace methods or referring to the realization of
the supersymmetry algebra in the Hilbert space of physical states.   

We have seen that in the renormalization of the one-loop self energies
and vertex corrections using DReg several non-invariant counterterms
are necessary. Still the calculation has been just as
straightforward as if we would have  relied on an invariant
regularization and used only invariant counterterms. The reason is
that the symmetry conditions may be used as an efficient tool for the
practical determination of counterterms. This is particularly
important for calculations beyond one-loop order since there the
behaviour of invariant but inconsistent schemes such as DRed
is not really under  control. One should note, however, that using
DRed in the 1-loop examples of this article invariant counterterms are
sufficient to renormalize correctly not only the self energies and vertex
corrections, as is well known \cite{CJN80}, but also the vertex
functions expressing the higher order corrections to supersymmetry
transformations. 

Higher order corrections to the non-linear supersymmetry
transformations are determined in terms of vertex functions involving
external $Y$ fields and $\epsilon$ ghosts and are in general
non-local. The corresponding counterterms may be read off from
appropriate symmetry conditions. As an example we have calculated the
one-loop corrections to the supersymmetry transformations of the
electron and selectron. Via the Slavnov--Taylor identity they appear
in the relations between physical vertex functions and may thus have
also phenomenological implications. 

The whole study can be generalized to supersymmetric models with soft
breakings and eventually to the supersymmetric extensions of the
standard model.  For the standard model the algebraic renormalization
has been worked out in \cite{Kraus97}, soft breakings have been
introduced in \cite{MPW96b}.  Although the corresponding
Slavnov--Taylor identities are more involved since they have to
express not only the symmetries but also the spontaneous or soft
breaking, their structure is the same as in  SQED. So it is possible
also for these models to derive symmetry conditions which may be
exploited in practical calculations if the existence of consistent
invariant regularization schemes is questionable. 


\begin{appendix}

\section{Conventions}

\subsection{Spinors}

\paragraph{2-Spinor indices and scalar products:}

\begin{eqnarray}\label{DefEpsilon}
\epsilon_{\alpha\beta} & = & - \epsilon_{\beta\alpha},\quad
\epsilon_{12} =1,
\quad\epsilon^{\alpha\beta}\epsilon_{\beta\gamma}  =  \delta^\alpha\ _\gamma
,\\
\epsilon_{\dot\alpha\dot\beta} & = & 
- \epsilon_{\dot\beta\dot\alpha},\quad 
\epsilon_{\dot1\dot2} =
1,\quad\epsilon^{\dot\alpha\dot\beta}\epsilon_{\dot\beta\dot\gamma} 
=  \delta^{\dot\alpha}\ _{\dot\gamma} 
,\\
\psi\chi & = & \psi^\alpha\chi_\alpha\ , 
\quad
\psi^\alpha  = 
\epsilon^{\alpha\beta}\psi_\beta\ ,
\\
\overline\psi\overline\chi & = & 
\overline\psi_{\dot\alpha}\overline\chi^{\dot\alpha}\ ,
\quad
\overline\psi_{\dot\alpha}  = 
\epsilon_{\dot\alpha\dot\beta}\overline\psi^{\dot\beta}\ .
\end{eqnarray}

\paragraph{$\sigma$ matrices:}

\begin{eqnarray}\label{Paulimatrizen}
&&
\sigma^1  = \left(\begin{array}{cc}0&1\\1&0\end{array}\right),\quad
\sigma^2  = \left(\begin{array}{cc}0&-i\\i&0\end{array}\right),\quad
\sigma^3  = \left(\begin{array}{cc}1&0\\0&-1\end{array}\right) ,
\\
&&
\sigma^\mu_{\alpha\dot\alpha}  =  (1, \sigma^k)_{\alpha\dot\alpha}
\ ,\quad 
\overline\sigma^{\mu\dot\alpha\alpha}  =  (1,
-\sigma^k)^{\dot\alpha\alpha}
\ ,
\\&&
\label{DefSigmaMunu}
(\sigma^{\mu\nu})_\alpha\ ^\beta  = 
\frac{i}{2}(\sigma^\mu\overline\sigma^\nu-\sigma^\nu\overline\sigma^\mu)
_\alpha\ ^\beta
\ ,\quad
(\overline\sigma^{\mu\nu})^{\dot\alpha}\ _{\dot\beta}  = 
\frac{i}{2}(\overline\sigma^\mu\sigma^\nu-\overline\sigma^\nu\sigma^\mu)
^{\dot\alpha}\ _{\dot\beta}\ .
\end{eqnarray}

\paragraph{Complex conjugation:}

\begin{eqnarray}
(\psi\theta)^\dagger & = & \thetabar\psibar \ ,\\
(\psi\sigma^\mu\thetabar)^\dagger & = & \theta\sigma^\mu\psibar\ ,\\
(\psi\sigma^{\mu\nu}\theta)^\dagger & = & \thetabar\sigmabar^{\mu\nu}\psibar\ .
\end{eqnarray}

\paragraph{Derivatives:}

\begin{eqnarray}
\frac{\partial}{\partial\theta^\alpha} \theta^\beta 
& = & \delta_\alpha{}^\beta
\ ,\quad
\frac{\partial}{\partial\theta_\alpha} \theta_\beta 
 = 
\epsilon^{\alpha\gamma}\epsilon_{\beta\delta}\delta_\gamma{}^\delta
= -\delta_\beta{}^\alpha
\ ,\\
\frac{\partial}{\partial\thetabar_\alphadot} \thetabar_\betadot 
& = & \delta^\alphadot{}_\betadot
\ ,\quad
\frac{\partial}{\partial\thetabar^\alphadot} \thetabar^\betadot 
 = 
\epsilon_{\alphadot\gammadot}
\epsilon^{\betadot\deltadot}\delta^\gammadot{}_\deltadot
= -\delta^\betadot{}_{\alphadot}
\ .
\end{eqnarray}

\paragraph{4-Spinors:}

The general relations between a 4-spinor and derivatives with respect
to it are defined in such a way that 
$
\dfunc{}{\Psi} \Psi = 1,
\dfunc{}{\Psibar} \Psibar = 1:
$
\begin{eqnarray}
&&\Psi = {\psi_\alpha \choose \overline\chi^{\dot\alpha}}\ ,\quad
\overline\Psi = \left(\chi^\alpha \ \overline\psi_{\dot\alpha}\right)
\ ,\\
&&\dfunc{}{\Psi} = \left(-\dfunc{}{\psi_\alpha},
  -\dfunc{}{\chibar^\alphadot}\right)\ ,\quad
\dfunc{}{\Psibar} = {\dfunc{}{\chi^\alpha} \choose
  \dfunc{}{\psibar_\alphadot}}\ .
\end{eqnarray}

\paragraph{$\gamma$ matrices:}

\begin{eqnarray}
&&\gamma^\mu  =  \left(\begin{array}{cc} 0 & \sigma^\mu \\ \overline{\sigma}^\mu &
0 \end{array}\right),\quad
\gamma^5    =  \left(\begin{array}{cc} -1 & 0 \\ 0 &
1\end{array}\right),\quad
P_{L,R}  = \frac{1\mp\gamma^5}{2}.
\end{eqnarray}

\subsection{Vertex functions}
\label{AppVertexFcts}
Vertex functions with external $\chi_1,\chi_2,\ldots$ are defined as
\begin{eqnarray}
\Gamma_{\chi_1\chi_2\dots}(x_1,x_2,\dots) =
\frac{\delta\Gamma(\varphi_i'=\varphi_i=Y_i=0)}
 {\delta\chi_1(x_1)\delta\chi_2(x_2)\dots}
\ .
\end{eqnarray}
The $\chi_i$ may be any of the physical fields, ghosts, or
$Y$ fields. For $\chi_i$ being one of the global ghosts it is
understood that there is no corresponding $x_i$ argument, and that the
functional derivative reduces to a partial derivative.
 
The sign of the momenta in Fourier transforms is defined in such a way
that momenta are always diagrammatically incoming. The Fourier
transform of vertex functions thus involves the opposite sign for the
momenta, as compared to the fields:
\begin{eqnarray}
&&\chi(x) = \int\frac{d^4p}{(2\pi)^4}e^{-ipx}\chi(p)\ ,
\\
&&(2\pi\delta)^4(p_1+\dots)\Gamma_{\chi_1\dots}(p_1,\dots)
= \int d^4x_1 \dots e^{-i(p_1x_1+\dots)}
\Gamma_{\chi_1\dots}(x_1,\dots)\ .
\end{eqnarray}

\subsection{One-loop integrals}
\label{App1LInt}
We use the following one-loop two- and three-point
functions \cite{PaVe79}:
\begin{eqnarray}
B_{\{0,\mu,\mu\nu\}} & := & \int
        \frac{\{1,k_\mu,k_\mu k_\nu\}}{[k^2-m_0^2][(k+p_1)^2-m_1^2]}
\ ,\\
C_{\{0,\mu\}} & := & \int
        \frac{\{1,k_\mu\}}
             {[k^2-m_0^2][(k+p_1)^2-m_1^2][(k+p_1+p_2)^2-m_2^2]}
\end{eqnarray}
with
\begin{eqnarray}
\int & \to & \mu^{4-D}\frac{16\pi^2}{i}\int \frac{d^Dk}{(2\pi)^D}
\end{eqnarray}
and the tensor decomposition
\begin{eqnarray}
B_\mu & = & {p_1}_\mu B_1
\ ,\\
B_{\mu\nu} & = & {p_1}_\mu{p_1}_\nu B_{21} + g_{\mu\nu} B_{22}
\ ,\\
C_\mu & = & {p_1}_\mu C_{11} + {p_2}_\mu C_{12}
\ ,\\
B_{ij} & = & B_{ij}(m_0^2,m_1^2,p_1^2)
\ ,\\
C_{ij} & = & C_{ij}(m_0^2,m_1^2,m_2^2,p_1^2,p_2^2,(p_1+p_2)^2)
\end{eqnarray}
in the conventions of \cite{aaff}.


\end{appendix}


\end{document}